\documentclass[CRPHYS,Unicode,manuscript]{cedram}

\renewcommand{\vec}{\boldsymbol}
\usepackage{overpic}
\usepackage{sistyle}
\usepackage{layout}

\usepackage{pdflscape}
\usepackage{amsmath, amssymb}
\usepackage{array}
\usepackage{booktabs}
\usepackage{multirow}
\usepackage{hyphenat}

\newcolumntype{M}{>{$}c<{$}}
\newcolumntype{L}{>{$}l<{$}}
\newcolumntype{R}{>{$}r<{$}}

\usepackage{tikz}
\usetikzlibrary{decorations.fractals}
\usetikzlibrary{calc,fit,spy}
\usetikzlibrary{mindmap,trees}
\usetikzlibrary{decorations.fractals}
\usetikzlibrary{shapes,arrows}
\usetikzlibrary{arrows.meta,decorations.markings,math,arrows.meta}
\usetikzlibrary{patterns.meta,patterns}
\usetikzlibrary{decorations.text}
\usetikzlibrary{positioning}
\usetikzlibrary{patterns,angles,quotes}
\usetikzlibrary{shapes.misc}
\usetikzlibrary{backgrounds}

\usetikzlibrary{chains,positioning,shapes.symbols,fadings,shadows, backgrounds}
\usetikzlibrary{decorations.pathmorphing}
\usetikzlibrary{shapes.callouts}
\usetikzlibrary{shapes.arrows, shadings}
\usetikzlibrary{decorations.text}

\newcommand{\Rt}{\widetilde{Ra}}

\definecolor{cublue}{HTML}{0066CC}
\definecolor{sodium}{RGB}{235, 235,235}
\definecolor{mercury}{RGB}{210,210,210}
\definecolor{gallium}{RGB}{180,180,180}
\definecolor{sulfur}{RGB}{241, 221, 56}
\definecolor{helium}{RGB}{224,255,254}
\definecolor{nitrogen}{RGB}{192,237,255}
\definecolor{sf6}{RGB}{162,210,223}
\definecolor{water}{RGB}{62,164,240}
\definecolor{sucrose}{RGB}{255,133,199}
\definecolor{novec}{RGB}{255,191,226}
\definecolor{acid}{RGB}{143, 254, 9}
\definecolor{silicone}{rgb}{1.0, 0.98, 0.98}
\definecolor{nonanol}{rgb}{1.0, 0.97, 0.0}
\definecolor{ethylene}{rgb}{0.86, 0.82, 1.0}
\definecolor{glycerol}{RGB}{186,85,255}

\definecolor{mulberry}{rgb}{0.77, 0.29, 0.55}
\definecolor{mediumseagreen}{rgb}{0.24, 0.7, 0.44}
 \definecolor{rbctop}{RGB}{0,170,255}
\definecolor{bluetop}{RGB}{0,124,186}
 \definecolor{rbcbottom}{RGB}{255,0,127}
 \definecolor{pinkbottom}{RGB}{255,0,127}

\tikzstyle{axial} = [fill = pinkbottom, opacity = 0.12, postaction={draw=pinkbottom!21,opacity = 1}]
\tikzstyle{axarrow} = [color = pinkbottom!18, line width= 5pt]

\tikzstyle{radial} = [fill = bluetop, opacity = 0.1,  postaction={draw =bluetop!50,opacity = 1}]
\tikzstyle{central} = [fill = mediumseagreen, opacity = 0.2,postaction={draw=mediumseagreen!60,opacity = 1}]
\tikzstyle{tangent} = [fill = mulberry, opacity = 0.3,postaction={draw=mulberry!70,opacity = 1}]
\title{Seven decades of exploring planetary interiors with rotating convection experiments}

\alttitle{Sept d\'ecennies d'exploration des int\'erieurs de plan\`etes par les exp\'eriences de convection en rotation}

\author{\firstname{Alban} \lastname{Poth\'erat}\CDRorcid{0000-0001-8691-5241}\IsCorresp}
\address{Centre for Fluid and Complex Systems, Coventry University, Mile Lane, CV1 2NL Coventry, United Kingdom}
\thanks{A.P. is supported by EPSRC (grant EP/X010937/1) and the Leverhulme Trust (grant RPG-2017-366)}
\email[A. Poth\'erat]{alban.potherat@coventry.ac.uk}
\author{\firstname{Susanne} \lastname{Horn}\CDRorcid{0000-0002-7945-3250}}
\addressSameAs{1}{Centre for Fluid and Complex Systems, Coventry University, Mile Lane, CV1 2NL Coventry, United Kingdom}
\email[S. Horn]{susanne.horn@coventry.ac.uk}
\thanks{S.H. is supported by the UKRI Horizon Europe guarantee, selected by the European Research Council (ERC), grant no. EP/X034402/1 (MAGNADO). S.H. further received funding from EPSRC, grant no. EP/V047388/1.}

\ESM{Supplementary material for this article is supplied as a separate
archive available from  the journal's
website under
article's URL
or from the author.}
\CDRsupplementaryTwotypes{supplementary-material}{\cdrattach{all_experiments.pdf}}


\begin{abstract}
The interiors of all the planets in the solar system consist of layers, most of which are made out of fluids. When these layers are subject to superadiabatic temperature or compositional gradients, turbulent convection takes place that transports heat and momentum. In addition, planets are fast rotators. Thus, the key process that underpins planetary evolution, the existence of dynamo action or lack thereof, the observable flow patterns, and much more, is rotating convection. Because planetary interiors are remote and inaccessible to direct observation, experiments offer crucial, physically consistent models capable of guiding our understanding and complementing numerical simulations. If we can fully understand the fluid dynamics of the laboratory model, we may eventually fully understand the original.
Experimentally reproducing rotating thermal convection relevant to planetary interiors comes with very specific challenges, in particular, modelling the central gravity field of a planet that is parallel to the temperature gradient. Three distinct classes of experiments have been developed to tackle this challenge. One approach consists of using an alternative central force field such as the electric one. This comes with the caveat that these forces are typically weaker than gravity and require going to space. Another method entails rotating the device fast enough so that the centrifugal force exceeds and effectively supersedes Earth's gravity. This mimics the equatorial and lower latitude regions of a planet. Lastly, insight into the polar and higher latitude regions is gained by using the actual lab gravity aligned with the rotation axis.
These experiments have been continuously refined during the past seven decades. Here, we review their evolution, from the early days of visualising the onset patterns of convection, over central force field experiments in spacecraft, ultrasound velocity measurements in liquid metals, to the latest optical velocity mapping of rotating magnetoconvection in sulphuric acid inside high-field magnets. We show how innovative experimental design coupled with emerging experimental techniques has advanced our understanding of planetary interiors and helped us paint a more realistic, detailed picture of them, including Earth's liquid metal outer core.
\end{abstract}

\begin{altabstract}
Les int\'erieurs de toutes les plan\`etes du syst\`eme solaire sont constitu\'es de couches
 internes, dont la plupart sont fluides. Quand ces couches sont sujettes \`a un gradient de temp\'erature ou de composition chimique super-adiabatiques, un \'ecoulement convectif 
s'installe qui transporte chaleur et moment angulaire. De plus, les plan\`etes tournent 
extr\^ement vite. Ainsi, le processus-cl\'e qui sous-tend l'\'evolution des plan\`etes est 
la convection en rotation. Alors que les int\'erieurs de plan\`etes sont eloign\'es et 
impropres \`a l'observation directe, les exp\'eriences de laboratoire pr\'esentent des mod\`eles coh\'erents \`a m\^eme de guider notre compr\'ehension et de compl\'ementer les simulations num\'eriques. Si nous parvenons \`a comprendre la m\'ecanique des fluides des mod\`eles de laboratoire, nous finirons peut-\^etre par comprendre l'original compl\`etement.
Cependant, reproduire exp\'erimentalement une convection en rotation pertinente pour les int\'erieurs de plan\`etes pr\'esente une gageure en soi, 
ne serait-ce qu'au niveau de la mod\'elisation de la gravit\'e centrale d'une plan\`ete et de son gradient de temp\'erature parall\`ele. Trois classes d'exp\'eriences distinctes sont apparues pour relever ce d\'efi. Une premi\`ere approche consiste \`a utiliser une autre force centrale, telle que la force de Coulomb. L'inconv\'enient est que ces forces sont typiquement plus faibles que la gravit\'e ambiante et n\'ecessitent donc d'aller dans l'espace. Une autre m\'ethode implique de faire tourner le dispositif suffisamment vite pour que la force centrifuge d\'epasse et m\^eme remplace la gravit\'e terrestre. Cette technique imite les conditions des regions \'equatoriales et \`a faible latitude d'une plan\`ete. Finalement,  utiliser la v\'eritable gravit\'e terrestre align\'ee avec l'axe de rotation nous renseigne sur les r\'egions polaires et \`a haute latitude.
Ces exp\'eriences ont \'et\'e perfectionn\'ees au cours de plus de sept d\'ecenies. Nous parcourons ici leur \'evolution, depuis les premi\`eres visualisations des motifs au d\'epart de la convection, en passant par les exp\'eriences \`a gravit\'e centrale conduites \`a bord d'engins spatiaux, l'av\`enement de la v\'elocim\'etrie \`a ultrasons dans le m\'etaux liquides, jusqu'aux derni\`eres cartographies de champs de vitesse par m\'ethodes optiques dans des \'ecoulements de magnoconvection en rotation dans de l'acide sulfurique \`a l'int\'erieur d'aimants \`a champs fort. Nous montrons comment des concepts exp\'erimentaux innovants coupl\'es aux techniques exp\'erimentales \'emergentes ont pouss\'e notre compr\'ehension des int\'erieurs de plan\`etes et nous ont aid\'e \`a  en d\'epeindre une image toujours plus d\'etaill\'ee et plus r\'ealiste, en particulier du c\oe ur liquide de la Terre. 
\end{altabstract}
\altkeywords{Convection en rotation, m\'ecanique des fluides expr\'erimentale, techniques de mesure, int\'erieurs de plan\`etes, convection turbulente}

\begin{document}

\maketitle

\section{Introduction}
In 1906, Richard D. Oldham \cite{Oldham1906}, wrote \emph{``Of all regions of the earth none invites speculation more than that which lies beneath our feet, and in none is speculation more dangerous: yet, apart from speculation, it is little that we can say regarding the constitution of the interior of the earth. [\ldots] The central substance of the earth has been supposed to be fiery, fluid, solid, and gaseous in turn, till geologists have turned in despair from the subject, and become inclined to confine their attention to the outermost crust of the earth, leaving its centre as a playground for mathematicians''}. Then the seismograph changed geophysics forever. Oldham analysed earthquake data and was indeed able to see inside the Earth. He concluded that the Earth had a mantle and core based on different travelling behaviours of seismic waves. In 1936, Inge Lehmann further proved seismologically that there was an inner solid core \cite{Lehmann1936}.

Seismology provides us with an accurate measure of the core-mantle and inner core boundaries (CMB, ICB) and unequivocal evidence that the outer core is liquid. Recent advanced seismological techniques have been revealing the detailed structure of Earth's mantle and inner core. They also complement other approaches, e.g. numerical simulations, high-pressure experiments or mineral physics modelling \cite{Billen2008, Olugboji2013, Mukhopadhyay2019, Long2013, Stixrude2012, Brodholt2000}. The outer core is a very different story. Its liquid nature makes it less amenable to seismology-based constraints. The main reason lies in the outer core's comparatively low viscosity and subsequent lack of shear strength. The outer core has a viscosity of about $\nu_{oc} \approx \SI{10^{-6}}{m^2/s}$, which is about $10^{24}$ and $10^{16}$ times lower than those of the mantle \cite{Normand1977} and the inner core \cite{Jeanloz1988}, respectively.

The low viscosity of Earth's outer core has another crucial consequence: Rotation matters. The relative importance of rotation is expressed through the Ekman number, $Ek = \nu/(2 \Omega H^2)$, where $\nu$ is the viscosity, $H$ is the layer thickness, and $\Omega$ is the rotation rate. Earth's rotation rate is about $\Omega_{E} = 7.27 \times \SI{10^{-5}}{rad/s}$, and the mantle, outer core and inner core have respective thicknesses of $H_{m} = \SI{2885}{km}$, $H_{oc} = \SI{2270}{km}$, and $H_{ic} = \SI{1216}{km}$. This gives $Ek_{m} \approx 10^9$, $Ek_{oc} \approx 10^{-15}$ and $Ek_{ic} \approx 47$, respectively. Throughout the entire interior of the Earth, convection drives motion and transports heat and momentum. But the high $Ek$ in the mantle and inner core means convection in those two regions is practically unaffected by rotation, whereas the low $Ek$ in the outer core implies that convection there is strongly constrained by rotation \cite{kunnen2021_jot,Ecke2023, aurnou2015_pepi, Aurnou2020}.
Another distinct feature of the outer core is that it is both made of electrically conducting metal and liquid. Because of it, convection is vigorous enough to power dynamo action and so generates and sustains Earth's magnetic field.

In short, the outer core is dynamically very different compared to the mantle or the inner core. It is the layer of the Earth that is least accessible through conventional means, in particular, seismology. Hence, the outer core is still largely a \emph{``playground''} for mathematicians, numericists, and, importantly, experimentalists. This by no means implies that it is all mere speculation or that we should not study the problem. Instead, to account for the large uncertainties on the actual conditions in Earth's interior, we need to create models that are rigorously built upon the known principles of mathematics, physics, and most crucially fluid dynamics. But as Willis puts it \cite{willis1894}, \emph{"[\ldots] it is less difficult to imitate one of nature's processes than to understand either the imitation or, through it, the original"}. In the study of outer core dynamics, it is experiments that most reliably ensure that our understanding is built on firm ground, and shall be the focus of this review.

We first discuss briefly the theoretical framework that is used to understand both the laboratory experiments and the planetary core processes these experiments aim to \emph{``imitate''}. Then, we give an overview of the major design principles and their historical development. In the main part, we review individual experiments, the challenges involved in experimental endeavours and the engineering strategy to tackle them. We shall also see how boldly extrapolating these models to the Earth as well as other planets encourages further exploration along this path while flagging up its limitations at the same time.

\subsection{Governing equations and control parameters}
To understand the challenges in bringing planetary physics to life in a small contraption at a laboratory scale, we shall start by summarising the main theoretical ingredients that both are usually assumed to share, i.e. rotating thermal convection. The most common and simplest framework describing rotating convection experiments exploring planetary interiors is that of the Navier-Stokes equations complemented by the transport equation for temperature $T$. The basic configuration is the Rayleigh-B\'enard convection (RBC) one \cite{benard1900_rgspa,rayleigh1916}, made of a fluid layer, heated from below and cooled from above (in the sense of gravity) with a superadiabatic temperature gradient. 
The fluid in planetary cores modelled by lab experiments is usually assumed to be Newtonian and to be described within the Oberbeck-Boussinesq approximation \cite{Oberbeck1879, Boussinesq1903, Gray1976}. Thus, its material properties, kinematic viscosity $\nu$ and thermal diffusivity $\kappa$ are homogeneous, temperature- and pressure-independent. The density is approximated as a Taylor expansion around $T = T_m$ up to the first order, i.e. it is assumed to be linearly dependent on the temperature, 
\begin{equation}
\rho = \rho_m (1 - \alpha (T-T_m)),
\label{eq:rho_oberbou}
\end{equation}
with $\alpha$ being the constant isobaric expansion coefficient, $\alpha \equiv - \frac{1}{\rho}\partial_T \rho \big\vert{}_p $. The subscript $m$ denotes the reference value, typically, the arithmetic mean temperature between the top and bottom boundary temperatures,  $T_m = (T_t+T_b)/2$, and $\rho_m = \rho(T_m)$. The second term in the density is small except in the gravitational and centrifugal buoyancy terms \cite{Marques2007, Horn2018, Horn2021, Horn2019, busse1970_jfm}. The temperature is the actual temperature in experiments and the superadiabatic part of the temperature when considering planetary interiors \cite{Busse2002b}. Planetary rotation is assumed constant, $\Omega\hat{\vec{e}}_z$, and the origin is placed at the centre of the planet.
The governing equations in dimensional form under these assumptions are:
\begin{eqnarray}
\vec{\nabla} \cdot \vec{u} &=& 0, \label{eq:cont_dim}\\
D_t {\vec{u}} &=&  \nu \vec{\nabla}^2 {\vec{u}} -  \vec{\nabla}p  + 2 \Omega  \vec{u}  \times \hat{\vec{e}}_z - \Omega^2 r \alpha (T-T_m) \hat{\vec{e}}_r + g \alpha (T -T_m) \hat{\vec{e}}_g, \label{eq:ns_dim}\\
D_t T &=&  \kappa \vec{\nabla}^2 {T}. \label{eq:energy_dim}
\end{eqnarray}
The unit vector $\hat{\vec{e}}_g$ points in the opposite direction of gravity, e.g. opposite to the spherical radial direction in spherical geometries or in the vertical direction in plane geometries, such as cylinders or annuli; $\hat{\vec{e}}_r$ is the cylindrical radial unit vector perpendicular to the direction of rotation $\hat{\vec{e}}_z$.

These equations are more meaningful in nondimensional form. Choosing as reference quantities, the fluid layer thickness $H$, the temperature difference between the hot and cold boundary, in the case of Earth, the inner core boundary (ICB) and core-mantle boundary (CMB), $\Delta$, the free-fall velocity $U_{f\!f}= \sqrt{g\alpha H\Delta }$, and the corresponding time and pressure scales, $H/U_{f\!f}$ and $\rho_m U_{f\!f}^2$, we have:
\begin{eqnarray}
\vec{\nabla} \cdot {\vec{u}} &=& 0, \label{eq:cont}\\
D_{{t}} {\vec{u}} &=& \sqrt{ \frac{Pr}{Ra}} \, \vec{\nabla}^2 {\vec{u}}  - \vec{\nabla} {p}  + \sqrt{\frac{Pr}{ Ra \, Ek^2}} \, {\vec{u}} \times {\hat{\vec{e}}_z} {- \frac{Fr}{\gamma} \, T \, r \, \hat{\vec{e}}_r} + {T} {\hat{\vec{e}}}_z, \label{eq:ns}\\
D_{{t}} {T} &=&   \displaystyle  \sqrt{ \frac{1}{ Ra \, Pr}} \, \vec{\nabla}^2 {T}.
\label{eq:energy}
\end{eqnarray}
In this form, the problem is governed by the following nondimensional control parameters: the Ekman number, which expresses the ratio of viscous to Coriolis forces, the  Rayleigh number, which is the ratio of buoyancy to viscous forces, the Prandtl number, which is the ratio of viscous to thermal diffusivities, the Froude number, which is the ratio of centrifugal to gravitational forces, and the radius-to-height aspect ratio (or alternatively, the diameter-to-height aspect ratio)
\begin{equation}
 Ek= \frac{\nu}{2 \Omega H^2}, \quad Ra= \frac{\alpha g \Delta H^3}{\nu\kappa}, \quad  Fr= \frac{\Omega^2 R}{g}, \quad  \gamma = \frac{R}{H}, \quad  \Gamma = \frac{2R}{H}.
\end{equation}
The form of eqs. (\ref{eq:cont}-\ref{eq:energy}) brings out three further classical parameters, the free-fall-based Reynolds, Rossby, and P\'eclet numbers,
\begin{equation}
 Re=\sqrt{\frac{Ra}{Pr}}, \quad Ro=\sqrt{\frac{Ra Ek^2}{Pr}}, \quad Pe=\sqrt{Ra Pr}.
\end{equation}
This Rossby number is often referred to as the convective Rossby number to distinguish it from the output parameter. Similarly, also the Reynolds and  P\'eclet numbers can be output parameters when defined with measured velocities \cite{Aurnou2020}.

\subsection{Main physical processes \label{sec:phys}}
The eqs.~\eqref{eq:ns_dim} and \eqref{eq:ns} include both the centrifugal and the standard gravitational buoyancy terms. Both act in a very similar way, that is, denser (colder) fluid moves in the direction of $\hat{\vec{e}}_r$ and $-\hat{\vec{e}}_g$, and lighter (warmer) fluid moves in the direction of $-\hat{\vec{e}}_r$ and  $\hat{\vec{e}}_g$, respectively. Specifically, in a plane geometry, cold fluid moves radially outwards and downwards and warm fluid moves radially inwards and upwards.
The Froude number expresses which of the buoyancy terms is dominant \cite{Horn2018, Horn2019, Horn2021, busse1970_jfm, Chandrasekhar1961}. In experiments, $Fr$ is always non-zero and in some experiments, it is the design principle that $Fr\gg1$ such that the centrifugal acceleration supersedes the gravitational one and so becomes the effective gravity (see section~\ref{sec:cyl}). 

{In most theoretical and numerical studies, $Fr$ is set to zero. This is justified in most planetary settings where centrifugal buoyancy is negligible at the largest scale, e.g. for Earth's outer core $Fr\simeq1.72\times10^{-3}$.}
In this case, the main effect of the Coriolis force on convective and other flows can be seen by considering the limit of fast rotation $Ek\rightarrow0, Ro\rightarrow0$ 
in the governing equations: away from boundary layers, the main balance is \emph{geostrophic}, i.e. between the Coriolis force and the pressure gradient, so the curl of eq. \eqref{eq:ns} imposes that
\begin{eqnarray}
\partial_z \vec{u}&=&\mathcal O(Ek,Ro)\rightarrow0,\\
\nabla_{r\phi}\cdot\vec{u}&=&\mathcal O(Ek,Ro)\rightarrow0,
\end{eqnarray}
where $\phi$ is the azimuthal coordinate and $\nabla_{r\phi}$ is the two-dimensional divergence in the $r$-$\phi$ plane. Hence, at the leading order, the Coriolis force makes the flow quasi-two-dimensional and quasi-horizontally solenoidal, a result known as the \emph{Taylor-Proudman  Constraint (TPC)} \cite{thomson1887_pmjs,proudman1916_prsa,greenspan1969}. It implies that a cylindrical fluid parcel must conserve its height through motion, or equivalently, that the flow must follow surfaces of constant height along the rotation direction, called \emph{geostrophic surfaces}. Since the TPC restricts possible fluid motions, the onset of stationary convection occurs at higher critical Rayleigh numbers $Ra_c$ than without rotation. In a plane Rayleigh-B\'enard configuration, $Ra_c \sim  Ek^{-4/3}$, and the non-rotating convection cells at onset give way to much thinner spiral cells, of length scale $\ell_c \sim Ek^{1/3}$ \cite{Chandrasekhar1961}. Importantly, in low Prandtl number fluids ($Pr < 0.68$) such as liquid metals, the onset modes are oscillatory \cite{Chandrasekhar1961, clune1993_pre, goldstein1993_jfm, Goldstein1994, Horn2017, Aurnou2018, Vogt2021}. The critical Rayleigh number and length scales are then $Pr$-dependent, namely, $Ra_c \sim (Ek/Pr)^{-4/3}$ and $\ell_c \sim (Ek/Pr)^{1/3}$. {These plane layer onset scalings hold asymptotically also in spherical and annular geometries \cite{roberts1968_prsa, busse1970_jfm, Barik2023, Dormy2004}.}

At higher criticality, the {TPC loosens, and the} cellular structures turn into convective Taylor columns, and then plumes and geostrophic turbulence (see e.g. the recent reviews by Kunnen \cite{kunnen2021_jot} and Ecke \& Shishkina \cite{ecke2023_arfm}).
{More generally, whenever inertia antagonises the effect of rotation, i.e. when the Rossby number based on the measured velocity $Ro_U$ is finite, the TPC is partly violated and flow structures become anisotropic with a ratio of their axial to transverse length scale governed by the balance between Coriolis and inertia forces $l_z/l_\perp\sim Ro_U^{-1/3}$ \cite{brons2020_jfm1,brons2020_jfm2,billant2021_jfm}, see the reviews by Godeferd \& Moisy \cite{godeferd2015_amr} and Biferale \cite{biferale2021_jot} on how anisotropy determines rotating turbulence.} 

{The Coriolis force may, however, have another important effect. It can lead, at least locally, to a misalignment of the time-averaged temperature gradient and gravity and therefore the main pressure gradient.
The misalignment drives a baroclinic flow (see the review by Harlander within this Special Issue \cite{harlander2024_crphys}). 
For example, in Earth's core outside the polar region, baroclinic flows naturally arise as the TPC aligns the flow with rotation rather than gravity \cite{aujogue2018_jfm, agrawal2024_gji}.
Similarly, in an experimental context or on a local planetary scale when $Fr > 0$, there is a flow even at arbitrarily low imposed temperature gradients.
The} Coriolis force constrains the baroclinic motion into an azimuthal wind $u_\phi$ which is controlled by the \emph{thermal wind balance}.
In cylindrical coordinates, and using the previous non-dimensionalisation, it reads
\begin{equation}
 \partial_z u_\phi = Ro \left(\partial_r T + \frac{Fr}{\gamma} r \partial_z T\right).\label{eq:tw_balance}
	\end{equation}
{and derives from the curl of eq.~\eqref{eq:ns_dim}.
The two examples mentioned above are illustrations of the first and second terms on the RHS of eq. \eqref{eq:tw_balance}, respectively, being significant.}
\subsection{Planetary and experimental peculiarities}
\begin{figure}[hpt!]
%
\includegraphics[width=\textwidth]{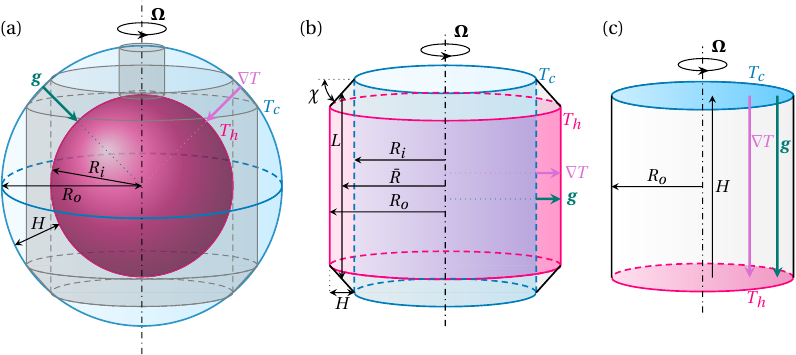}
	\caption{Sketches of the fundamental geometries used for the study of rotating convection in planetary core convection. The gravitational acceleration (or any proxy) $\vec{g}$ and the temperature gradient $\vec{\nabla} T$ and their respective directions are indicated by dark green and lavender arrows. The angular rotation is assumed to be vertical,  $\vec{\Omega} = \Omega \hat{e}_z$, and the axis of rotation is marked by the vertical black dash-dotted line. The boundaries are demarcated by blue for the cold boundary $T_c$ and pink for the hot boundary $T_h$. (a) Spherical shell;  $R_i$ is the inner radius, $R_o$ is the outer radius, and $H = R_o - R_i$ is the spherical shell gap width. The grey cylinder of radius $R_i$ is the tangent cylinder (TC). The grey annulus is detailed in (b), and the small grey cylinder in the polar region is detailed in (c). (b)~\emph{Busse annulus}; $R_i$ is the inner radius, $R_o$ is the outer radius, $H = R_o - R_i$ is the annulus gap width, $\bar{R} = R_i + H/2 = R_o - H/2$ is the mean radius, $L$ is the mean height, and $\chi$ is the sloping angle of the endwalls. The sidewalls are typically thermally insulating. (c) Cylinder; $R_o$ is the radius, $H$ is the height of the cylinder. The endwalls are typically thermally insulating. \label{fig:geometries}}
\end{figure}
In the planetary context, the spherical shell geometry constrains the expression of these processes. 
Since planets are extremely fast rotators, 
the flow tends to follow geostrophic contours because of the TPC. Spherical shells being axisymmetric, these geostrophic contours are cylinders aligned with the rotation axis, so radial motions are impeded whereas azimuthal flows are favoured by rotation. One of these contours plays a particularly important role: the so-called \emph{tangent cylinder (TC)} extruded from the equatorial boundary of the inner solid core along the rotation separates polar regions located below the poles from equatorial regions (see figure \ref{fig:geometries}a). If a fluid parcel was to cross the TC, it would see either a doubling or a halving of the domain's height along $\hat{\boldsymbol e}_z$, which the TPC strongly opposes. Hence the TC acts as a mechanical boundary between the polar and equatorial regions. For the same reason, radial motion near the equatorial plane is very strongly suppressed by the jump in domain height there.
Thus, because of the Earth's solid inner core, the liquid outer core is split into two regions with different types of rotating convection \cite{Gastine2023,agrawal2024_gji}.

In the equatorial region outside the TC, the gravity is rather perpendicular to the rotation. Near the equatorial plane, the gravity and temperature gradients drive motion in the plane, and therefore perpendicular to $\Omega\hat{\boldsymbol e}_z$. This motion is therefore not as strongly opposed by the TPC.
Instead the rotation tends to elongate these cells along $\hat{\boldsymbol e}_z$ and so forms quasi-geostrophic columns extending across the entire northern and southern hemispheres, named \emph{Busse columns}, after Busse's pioneering experimental work with the \emph{Busse annulus} represented in figure \ref{fig:geometries}(b) \cite{busse1976_sci}.
\begin{figure}
\centering
\includegraphics[width=\textwidth]{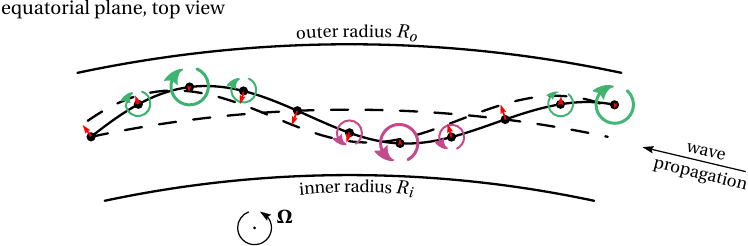}
\caption{Sketch of the general propagating behaviour of (incompressible) Rossby waves, adapted from ref. \cite{Busse2009}. Visualised is part of the equatorial plane of a rotating spherical shell.
Initially, a series of fluid columns is at rest and aligned along the mid-radius. If these columns are displaced in a sinusoidal manner, then the vortices that move outwards  (indicated by the green arrows) must reduce in length to conserve mass, thus, they get squashed and acquire retrograde vorticity to conserve potential vorticity. Similarly, the vortices that move inwards (indicated by the purple arrows) must extend in length, they get stretched and acquire prograde vorticity. The effect on the neighbouring columns is that the columns to the west are pushed away from the rotation axis and the columns to the east are pushed towards the rotation axis (red arrows). This results in the phase of the initial sinusoidal pattern propagating westwards (prograde) which is known as the Rossby wave.\label{fig:Rossbywaves}}
\end{figure}
The variation in domain height in the equatorial region and in the Busse annulus 
incurs a linear damping 
on the geostrophic structures 
called $\beta$-effect \cite{greenspan1969,pedlosky1987}. The reduced domain height at low latitudes prevents Busse columns from entering that region so they tend to remain close to the inner core.  
Nevertheless, the $\beta-$effect incurred by the sloping outer boundary forces Busse columns to oscillate radially and form \emph{thermal Rossby waves}, as sketched in figure \ref{fig:Rossbywaves}. Hindman \& Jain \cite{Hindman2022} offer an elegant explanation of their basic 
mechanism: Considering a column of height $L$, 
\emph{"If a spinning column near the equator is pushed toward the rotation axis, the column grows in height as the chord length of the column’s axis increases. In an incompressible fluid, this vortex stretching is accompanied by a commensurate narrowing of the column to conserve mass. Subsequently, as the column compresses laterally, the column must spin faster to conserve angular momentum about its own axis \cite{hide1966_prsa,busse1970_jfm}.
[..]  This conservation principle is enforced by the constancy of the potential vorticity. As $L$ increases,} [the vorticity component aligned with the rotation axis] \emph{ must also increase. The resulting induced vorticity causes the neighbouring
column to the west to be pushed outward, away from the rotation axis, and the column to the east to be pushed inward, toward the rotation axis. These newly pushed columns conserve their own potential vorticity (i.e., angular momentum) and induce spinning columns further down the belt to also move inward and outward. The result is a prograde-propagating
Rossby wave where the spinning columns dance back and
forth, toward and away from the rotation axis."} 
One of the main design challenges in imitating this phenomenology 
arises out of the vertical direction of gravity in the lab, which makes it difficult to reproduce equatorial convection in experiments using Earth's gravity. Here the centrifugal forces 
comes to the experimentalist's rescue. 
Since the region outside the TC is annular, it can be simulated by rotating an annulus with sloping endwalls sufficiently fast for the centrifugal force to exert an apparent radial gravity. This was the essence of Busse's revolutionary idea \cite{busse1974_jfm}.

In the polar regions inside the TC, the temperature gradient, the gravity and the rotation are mostly aligned, and topographic effects are minor: the convection there is expected to develop somewhat similarly to rotating RBC (RRBC). Experimentally, one would expect to capture its main features by rotating a vertical cylindrical vessel around its axis at a sufficiently large speed for the Coriolis force to dominate, as in figure \ref{fig:geometries}(c). Unfortunately, at the scale of the laboratory, the centrifugal force becomes important at rapid rotation, and even more so for cylinders of low aspect ratios $\Gamma=2R_o/H$ \cite{Horn2018}. Such a cylinder may represent the entire TC or a smaller, concentric cylinder inside it, see figure \ref{fig:geometries}(a,c). 
A further issue arises from the nature of the TC: while it may be legitimate to model it with a rigid boundary, the thermal boundary condition there is both unknown and crucial: if adiabatic, isothermal surfaces remain horizontal all the way across the cylinder, whereas if isothermal, very strong baroclinicity drive a flow near the cylinder side wall\cite{aujogue2016_rsi,aujogue2018_jfm,agrawal2024_gji}. 
Since there is no reason to think that either of these ideal cases applies to planetary cores, actual TCs must be reproduced in experiments.

{Besides topographic effects, solid boundaries in planets and rotating experiments give rise to viscous boundary layers. Along boundaries perpendicular to the rotation axis, Ekman layers, of thickness $\mathcal{O}(Ek^{1/2})$ result from the balance between the viscous and the Coriolis force \cite{ekman1905_amaf,greenspan1969}. They become extremely thin in regimes of rotation relevant to planets, and fully resolving them curtails the range of Ekman numbers accessible to numerical simulations. One of their most distinctive property is that wall friction within the layer occurs at an angle with the bulk velocity. This results from the progressive deflection of the flow by the Coriolis force through the layer forming the so-called \emph{Ekman spiral}. The second remarkable property of Ekman layers is that local axial vorticity in the bulk $\omega_z$ pulls a local mass flux out of the layer $u_z\simeq\frac12\omega_zEk^{1/2}$. This secondary flow, called \emph{Ekman pumping} is driven by the imbalance between the radial pressure gradient induced in the Ekman layer by rotation in the bulk, and the reduced centrifugal force due to low velocities within the layer. This phenomenology even persists in the presence of planetary or laboratory magnetic fields but follows different scalings \cite{acheson1973_arfm,potherat2000_jfm,davidson2002_ejmb}.}

{Boundary layers along vertical boundaries are \emph{Stewartson layers} \cite{stewartson1957_jfm,greenspan1969}. They exhibit a sandwiched structure made of two layers. A layer of thickness $\mathcal{O}(E^{1/4})$ arises from the balance between the Coriolis force and viscous friction in planes perpendicular to the rotation, while a thinner $\mathcal{O}(E^{1/3})$ is driven by secondary flows in planes meridional to the rotation axis. Kunnen et al. \cite{kunnen2013_jfm} offer a detailed analysis of their structure in the context of rotating  Rayleigh-B\'enard convection.}

{Vertical boundaries are also associated with specific convective structures called \emph{wall modes}, with an onset at lower Rayleigh numbers than the bulk modes \cite{Zhong1991, Zhong1993, herrmann1993_jfm, kuo1993_pre, goldstein1993_jfm, Goldstein1994, zhang2009_jfm,Ecke2023b}. 
It is argued that their existence may be connected to that of boundary zonal flows \cite{Zhang2020, Zhang2021, Vogt2021,ecke2022_prf}. These modes are robust to changes in the shape of the vertical boundary \cite{deWit2023, Terrien2023} and since they compete with bulk modes relevant to planets, and efficiently carry heat, their presence complicates the interpretation of experimental results in geophysical context.}

Finally, a key question that arises when experimentally modelling either region is the choice of the working fluid, a choice that is closely linked to the envisaged measurement techniques and the specific aims pursued in building such devices. 
The Earth's outer core is mostly made of liquid iron, whose physical properties can be estimated from Iron's melting point \cite{aubert2001_pepi, stacey1992}, {and high-pressure experiments \cite{xu2021_epsl,morard2013_epsl,gonzalez2023_jgr,pozzo2012_nat}} $\rho\simeq \SI{10^4}{kg/m^3}$, $\nu\simeq7\times\SI{10^{-6}}{m^2/s}$, $\kappa\simeq4\times\SI{10^{-6}}{m^2/s}$, $\alpha\simeq$\SI{10^{-5}}{K^{-1}}, so $0.1\leq Pr\leq1$ \cite{schubert2011_pepi}. 
Liquid metals would therefore seem the obvious choice but have only been used in a handful of experiments.
The main issue with liquid metals besides their direct or indirect costs, toxicity, and chemical reactivity is that they are opaque and so preclude any direct optical visualisation or any of the convenient LASER-based techniques used in transparent fluids to map velocity and temperature fields. For these reasons, high $Pr$ fluids such as water, but also silicon oils and a few more exotic fluids, appear as cheap and attractive alternatives to understand the interplay 
between convection and rotation in planetary geometries. 
For the purpose of studying the effect of magnetic fields, however, liquid metals are usually the fluid of choice. However, neither liquid metals nor the high Prandtl fluids mentioned above afford the combination of electromagnetic effects with the advantage of optical visualisation.
Sulphuric acid recently emerged as a solution to do just that \cite{andreev2013_jfm,moudjed2020_ef,aujogue2016_rsi,potherat2023_prl}.\\
\subsection{Historical evolution through four classes of experiments}
\begin{figure}
\includegraphics[width=\textwidth]{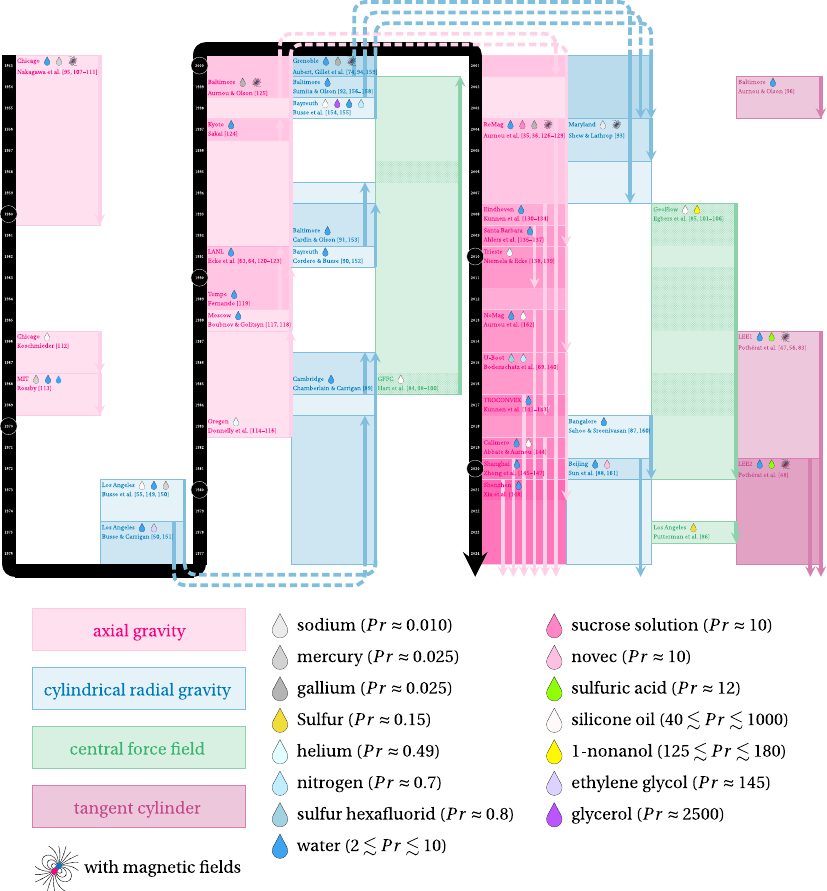}\\[1em]
  \caption{Timeline of the laboratory experiments relevant for planetary core convection from 1953 till 2024. As starting point the year of the date of submission of the first publication, or, where available, the date of the first experimental run is used, the endpoint is the year of the publication of the last publication of the experimental data. The experiments are either referred to by the name of the device or by the place they were conducted if the experimentalists were less creative with their naming convention than with the experimental design. The main experimentalist(s) are also given together with a selection of the relevant publications. The background colours indicate the class of experiment: pink - axial gravity, i.e. either cylinder or cuboid; blue - cylindrical radial gravity, i.e. cylindrical or spherical annulus, hemispherical or spherical shell; green - central force fields, i.e. sphere and hemisphere (the dotted areas indicate when the space missions took place); purple -  tangent cylinder geometry. The working fluids in the experiments are indicated by the different droplets. If rotating magnetoconvection experiments were performed, a magnet symbol is added. An A3 scale version of the figure is also provided as a supplementary file. \label{fig:timeline} 
}
\end{figure}

\begin{figure}
\centering
\begin{overpic}[width=\textwidth]{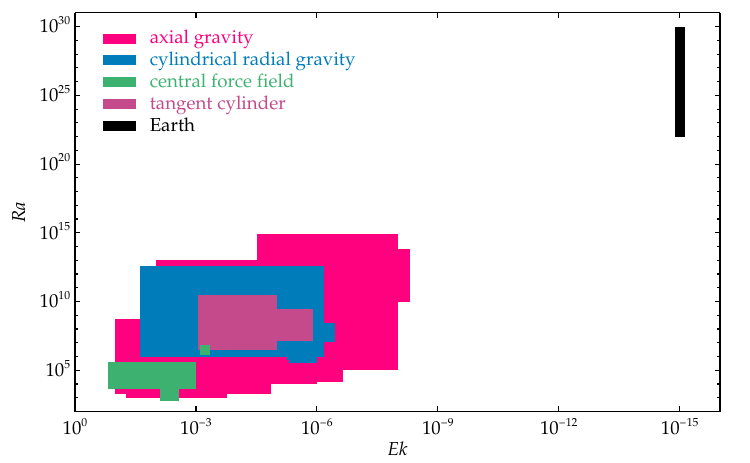}
\end{overpic}
\begin{overpic}[width=\textwidth]{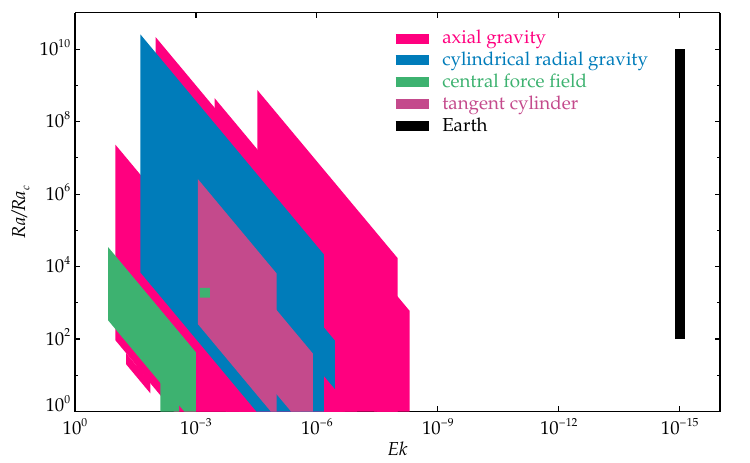}
\end{overpic}
\caption{Parameter space spanned by the four types of experiments modelling rotating convection in the Earth's outer liquid core, and parameter range where the actual Earth's outer core is currently believed to operate. While in $(Ra,Ek)$ space, all experiments sit very far from realistic geophysical regimes, especially, in terms of Rayleigh numbers, their levels of criticality $Ra/Ra_c$ may well reach the Earth's. The data represented here is given in tables~\ref{tab:dim_parameters} and \ref{tab:nondim_parameters}. \label{fig:parameters_space}}
\end{figure}

The constraints imposed by terrestrial gravity led to four fundamental classes of rotating convection experiments that explore planetary interiors.
However, with an outer core thickness of $H\simeq2.26\times\SI{10^6}{m}$, a rotation of one revolution per (Earth) day, a mid-thickness gravity of $g\simeq$\SI{7}{m/s^2} \cite{schubert2011_pepi},
the regimes of the outer core of Earth are too extreme for any laboratory experiment: $Ek\simeq 10^{-15}$, $Ra\simeq 10^{22}$ to $10^{30}$, $Ro\in[10^{-6},10^{-3}]$, $Re\in[10^{11},3\times10^{15}]$ and $Pe\in[10^{21},10^{31}]$. Other planets offer no solace either as they operate in just as extreme regimes \cite{schubert2011_pepi,aurnou2015_pepi}. 
Hence, experiments first focused on understanding the fundamentals of convection (onset, regimes, heat flux) to later extrapolate them to these extreme regimes, and incorporate some of the more complex specifics of the Earth (inhomogeneities, magnetic field).
A timeline summarising the history of about seven decades of experiments (1953--2024) is given in figure~\ref{fig:timeline}, and a large A3 version of it is provided as supplementary material. {Representations of the regimes of parameters in $(Ra,Ek)$ and in $(Ra/Ra_c,Ek)$ spaces spanned by all four types of experiments are shown in figure~\ref{fig:parameters_space} along with projected Earth's regimes. The detail of dimensional and non-dimensional parameters for experiments conducted by each team identified in the chronological diagram of figure~\ref{fig:timeline} is reported in tables~\ref{tab:dim_parameters} and \ref{tab:nondim_parameters}, respectively.} In the following, we give a brief overview of the four fundamental classes of rotating convection experiments.
\paragraph{Central force field} It is extremely hard to produce self-gravitating spheres in miniature form within a terrestrial laboratory environment, thus, proxy-gravitational forces have so far been the only way to generate central force fields similar to Earth's gravity and suitable for studying rotating convection. Examples are the dielectrophoretic force which was utilised in two space experiments (GFFC, 1985--1999 \cite{hart1986_sci} and GeoFlow, 2008--2020 \cite{egbers1999_mst}) and very recently the pycnoclinic acoustic radiation force which also works on Earth (Los Angeles, 2023 \cite{Koulakis2023}). These experiments are detailed in section~\ref{sec:central}.
\paragraph{Cylindrical radial gravity} This class of experiments simulates the regions outside the TC. Since gravity is mostly cylindrical-radial there, these experiments use annular vessels of various shapes: spherical, hemispherical, cylindrical and variations thereupon,  
detailed in section \ref{sec:cyl}.
Guided by these constraints, the first experiments were pioneered by Busse in the 1970s in a rotating annulus, with slanted endwalls reproducing the $\beta$-effect and focussed mostly on the onset of convection \cite{busse1974_jfm}. Simpler cylindrical annuli with flat endwalls were later used to study inhomogeneous heat flux\cite{sahoo2020_jfm} and ultimate convection \cite{jiang2020_sciadv}.
The first and only experiment in a full sphere in 1986 is due to Chamberlain \cite{chamberlain1986_gafd}. 
Aside of this, spherical annuli became the geometry of choice for groups in Los Angeles\cite{busse1976_sci}, Cambridge\cite{chamberlain1986_gafd}, Bayreuth\cite{cordero1992_grl}, Baltimore \cite{cardin1992_grl,sumita1999_sci}, Grenoble \cite{aubert2001_pepi} and Maryland \cite{shew2005_pepi}. 
The mid-2000s mark the appearance of experiments on rotating magneto-convection in spherical annuli in Grenoble \cite{gillet2007_jfm2} and Maryland \cite{shew2005_pepi}, using gallium and sodium, respectively.
\paragraph{Axial gravity} Experiments relevant to the polar region started {within} the convection community, rather than with geophysicists. 
The combination of axial gravity and rotation, with the geometry of the geostrophic contour, makes cylindrical vessels the obvious choice for this type of experiment. 
Indeed the very first experiment on rotating convection was conducted on the 20\textsuperscript{th} of November 1953 in Chicago by Fultz, Nakagawa, \& Frenzen \cite{Fultz1954}. Since then, cylinders have remained the experimentalists' favourite to study rotating convection with axial gravity. 
A very brief summary of the numerous experiments of this type is given in section~\ref{sec:axial}.
\paragraph{Tangent Cylinder Dynamics}  
The radial and axial strands met in 2003 when Aurnou et al \cite{aurnou2003_epsl} built the first experiment dedicated to studying the TC, with a raised heater inside a hemispherical dome filled with water. 
It wasn't until 2014 that this idea was pushed 
when Aujogue et al. used electrically conducting sulphuric acid and the availability of high magnetic fields up to 10 T to produce the study of rotating magneto-convection with PIV visualisations in a configuration directly relevant to planets \cite{aujogue2016_rsi,aujogue2018_jfm,aujogue2016_phd,potherat2023_prl}. This was followed by a much-improved rebuild using a cylinder instead of a hemisphere \cite{agrawal2024_gji} (see section~\ref{sec:axial}). 
%
%
\begin{table}[hpt!]
{
\renewcommand{\arraystretch}{1.05}
\setlength{\tabcolsep}{3.1pt}
\resizebox{\textwidth}{!}
{
\def~{\hphantom{-}}
\begin{tabular}{lMMMMMM}
\toprule
experiment &  \Omega [\mathrm{rad/s}] & \Delta [\mathrm{K}] & R_i [\mathrm{cm}] & R_o [\mathrm{cm}] &  H[\mathrm{cm}] \\[3pt]
\midrule
	&&&\multicolumn{2}{c}{Central force field}\\
Hart et al. (GFFC) \cite{hart1986_sci, hart1986_jfm, Hart1999, hart1996_3mg} &  [0.03,3.14] & [0.2,20] & 2.402 & 3.300 & 0.908  \\
Egbers et al. (GeoFlow) \cite{egbers1999_mst, egbers2003_asr, Futterer2010, Futterer2012, Futterer2013, Zaussinger2017, Zaussinger2020} & \{0.05,5,10\} & [0.1,10] & 1.35 & 2.70 & 1.35 \\ 
Putterman et al. \cite{Koulakis2023} & 106.8 & - & - & 3 & 3 \\ \midrule
	&&&\multicolumn{2}{c}{Axial gravity}\\

Nakagawa et al. \cite{Fultz1955, Nakagawa1955, Nakagawa1957, Nakagawa1959, Goroff1960}&0.52,1.05&n/a&-&7.25,10.5,12.5&3,5&\\
Nakagawa et al. \cite{Fultz1954}&[0.26,\pi]&n/a&-&7&[4,8]&\\

Koschmieder \cite{Koschmieder1967}&[\pi/2,2\pi]&[1.25,2.5]&-&10.5&3 \\
Rossby \cite{Rossby1969}&[0.1,4.83]&n/a&-&11.1125&0.5,1,2,3.5\\ 
Donnelly et al. \cite{Lucas1983, Pfotenhauer1984,Pfotenhauer1987}&[0.11, 1.981]&[2.1,3.6]&0.81,1.25&0.16,0.25\\
Boubnov \& Golitsyn  \cite{Boubnov1986, Boubnov1990}&[0.1,2\pi]&-&-&8.5&[2,24] \\
Fernando \cite{Fernando1991}&[0.07,1]&-&-&30 (\textrm{cube})&50\\
Ecke et al.  \cite{Liu1997, Zhong1991, Zhong1993, Ning1993, Liu2009, Vorobieff1998}&[\pi/50,4.9]&n/a&-&5,6.35,3.65(\textrm{square})&5,12.7,9.4(\textrm{square})\\
Sakai  \cite{Sakai1997}&[0.52,2.1]&n/a&-&10&3,6,9\\
Aurnou \& Olson  \cite{Aurnou2001}&[7.3\times10^{-3}, \pi]&n/a&-&7.6(\textrm{square)}&1.85,2.5\\
Aurnou et al. (RoMag) \cite{King2009, Cheng2015, Aurnou2018, King2013, Vogt2021, Grannan2022} & [0.7,3.1] & [0.4,60] & - & 20&  \{5,10,20,50\} \\
Kunnen et al. \cite{Kunnen2008, Kunnen2010, Rajaei2016,Joshi2016, Rajaei2017}&[0.01,4.1]&n/a&-&10,11.5,12.5&10,11.5,12.5\\
Ahlers et al. \cite{Zhong2009, Weiss2010, Weiss2011}&n/a&n/a&-&24.8&24.8,49.5 \\
Niemela \& Ecke \cite{Niemela2010,Ecke2014} & [0.03,0.2] & [0.03,0.2] & - & 50 & 100 \\
Niemela \& Ecke \cite{Niemela2010,Ecke2014} & [0.05,1.8] & [0.4, 25]& - &\{20,60\} & \{10,20,40,80,180\}  \\
Bodenschatz et al. (U-Boot) \cite{Zhang2020,Wedi2021} & [0.002,1.3] &[1,10.6] & - & 110 & \{110,220\}  \\
Kunnen et al. (TROCONVEX) \cite{cheng2018_gafd, Cheng2020,  Kunnen2021} & [0.15,2.2]& [1,25]& - & 40 &  \{80,200,300,400\}   \\
Abbate \& Aurnou ( Calimero) \cite{Abbate2023} &  [8.4 \times 10^{-2}, 3.7] & [2,55] & - & 9.64 & \{19.05, 38.10, 80.01\} \\
Zhong et al. \cite{Ding2021, Lu2021, Shi2020}&n/a&n/a&-&12&[6.3,60] \\
Xia et al. \cite{Hu2022}&[0.63,6.1]&[1,34]&-&9.8&9.8 \\ \midrule
	&&&\multicolumn{2}{c}{Cylindrical radial gravity}\\
	Busse et al. \cite{busse1974_jfm,Busse1982, Azouni1986} &n/a&n/a&n/a&n/a&n/a\\
Busse \& Carrigan \cite{busse1976_sci,carrigan1983_jfm} &n/a&n/a&n/a&n/a&n/a\\
Chamberlain \& Carrigan \cite{chamberlain1986_gafd} & [16,80]&7&0&7.51& 7.51 \\
Cordero \& Busse \cite{cordero1992_grl, Cordero1993} & 17.80,19.89&[0.7,3.7]&5&6.35,8.94&1.35,2.94 \\
Cardin \& Olson \cite{cardin1992_grl,cardin1994_pepi} &[26.2,41.9]&[0,10]&5&15&10 \\
Busse et al. \cite{Jaletzky2000, Westerburg2003} & n/a&n/a&n/a&n/a&n/a\\
Sumita \& Olson  \cite{sumita1999_sci,sumita2000_pepi,sumita2002_jgr,sumita2003_jfm} &21.57&[0,10]&5&15&10\\
Aubert et al. (Water) \cite{aubert2001_pepi} &[20.9,83.8]&[0,25]&4&11&7\\
Aubert et al. (Gallium) \cite{aubert2001_pepi} &[41.8,83.8]&[0,30]&4&11&7 \\
Gillet et al. (Water) \cite{gillet2007_jfm1}  &[20.9,83.8]&[8,33]&4&11&7 \\
Gillet et al. (Gallium)  \cite{gillet2007_jfm1, gillet2007_jfm2} &[20.9,83.8]&[8,33]&4&11&7 \\
Sahoo \& Sreenivasan \cite{sahoo2020_epsl, sahoo2020_jfm} & 31.31 &  -   & 0.051 & 0.142 & 0.092 \\
Sun et al. \cite{jiang2020_sciadv, jiang2022_prl}&[12,74]&n/a&12&12&12\\ 
\midrule
	&&&\multicolumn{2}{c}{Tangent cylinder}\\
Aurnou \& Olson \cite{aurnou2003_epsl}  &[4\times10^{-2},19.5]& - &5 &15.2& [11.6,15.2]& \\
Poth\'erat et al. (LEE1) \cite{aujogue2016_rsi, aujogue2018_jfm, potherat2023_prl} &[1.52, 6.28]&[0.3,25]&5&13.8, 14.25&12 \\
Poth\'erat et al. (LEE2) \cite{agrawal2024_gji} & 0,[0.75,6.28]&[0.3,23]&7.75&14.3&11 \\
\bottomrule\\
\end{tabular}}
\caption{Rotation rate $\Omega$, temperature difference $\Delta$, inner radius $R_i$ (heater radius in TC geometries), outer radius $R_o$ (corresponds to radius, i.e. half the diameter $D$ in cylinders, outer vessel radius in TC geometries), height $H$ (corresponds to gap width in spherical and other annulus experiments). 
No parameters are given in Busse \& Carrigan (1976). The parameters for the experiments by Putterman et al. \cite{Koulakis2023} are based on acoustic values.
\label{tab:dim_parameters}}
}
\end{table}

\begin{landscape}
{
\renewcommand{\arraystretch}{1.15}
\setlength{\tabcolsep}{3.1pt}
\resizebox{598pt}{!}
{
\def~{\hphantom{-}}
\begin{tabular}{lMMMMMMM}
\toprule
experiment &  Pr & Ra & Ek & Fr & Ro & \Gamma  \\[3pt]
\midrule
Hart et al. (GFFC) \cite{hart1986_sci, hart1986_jfm, Hart1999, hart1996_3mg} & 8.4 & [4.1\times10^3,4.3 \times 10^5] & [1.3\times 10^{-3}, 1.5\times10^{-1}] & [1.7\times10^{-5},0.26] &  [0.03,33] & 2.65 \\
Egbers et al. (GeoFlow) \cite{egbers1999_mst, egbers2003_asr, Futterer2010, Futterer2012, Futterer2013, Zaussinger2017, Zaussinger2020}  & \{64,125,178\} & [5.6\times10^2, 2.2 \times 10^5] & [2.6 \times 10^{-3}, 7.6 \times 10^{-3}] & [5.1\times 10^{-5}, 2.55] & [0.07,3.1] &  [17,75.0] \\
Putterman et al. \cite{Koulakis2023}  &  0.15 & 3 \times 10^6 & 6 \times 10^{-4} & 0.026 & 2.7 & -\\ \midrule
Nakagawa et al. (mercury) \cite{Fultz1954, Fultz1955, Nakagawa1957, Nakagawa1959, Goroff1960} &0.024 & [10^4,1.27\times10^{11}]&[10^{-6},3.46\times10^{-4}]&[0.36,8.1]\times10^{-3}&[0.13,1.2]&2.8, 2.9\\
Nakagawa et al. (water) \cite{Nakagawa1955}&6.4 & [10^5,9\times10^8]&[7\times10^{-6},10^{-3}]&[7\times10^{-3},1]&[6.24\times10^{-4},8.32\times10^{-2}]&[1.66, 15]\\
Koschmieder \cite{Koschmieder1967} &881&[9.9,20]\times10^2 & [1.34, 5.38]\times10^{-2}&[0.025,0.4] & [1.4,8.1]\times10^{-2}& 20.6\\
Rossby \cite{Rossby1969}&[0.025,200]&[2\times10^3,6\times10^6]&[1.6\times10^{-5},1.8\times10^{-2}]&[1.1\times10^{-4},0.26]&[5\times10^{-5},283]&6.4,11.1,22.2,44.5\\ 
Donnelly et al. \cite{Lucas1983, Pfotenhauer1984,Pfotenhauer1987} &[0.49,0.76] &[10^3,4\times10^6]& [0.03,1.7\times10^{-4}] &[9.2\times10^{-5},3.36] &[0.016,4.28] & 3.2,4.9,7.8\\
Ecke et al.  \cite{Liu1997, Zhong1991, Zhong1993, Ning1993, Liu2009, Vorobieff1998}&[3,7]&[2\times10^3,5\times10^8]&0,[0.1,1.4\times10^{-5}]&[1.1\times10^{-5}, 0.66]&[1.17\times10^{-4},8.9]& 1,2,5,0.78(\rm{square})\\
Sakai  \cite{Sakai1997} (square vessel)& 6.4&[10^6,10^9]&[1.4\times10^{-5},1.1\times10^{-3}]&[2.8\times10^{-1}, 0.1]&[1.1\times10^{-2},0.44]&2.2,3.3,6.7\\
Aurnou \& Olson  \cite{Aurnou2001}& 0.023&[1.5\times10^3, 2.9\times10^4]&[3.2\times10^{-3},3.2\times10^{-2}]&[4.15\times10^{-7}, 7.6\times10^{-2}]&[0.11, 8.1]&4 (\textrm{square}] \\
Aurnou et al. (RoMag) \cite{King2009, Cheng2015, Aurnou2018, King2013, Vogt2021, Grannan2022} & 0.025 & [1.4\times10^4, 2.1\times 10^9]& [ 2.3 \times 10^{-7}, 1 \times 10^{-4}] & [10^{-2},0.2] & [1.7\times 10^{-4}, 3.0]&  \{0.4,1,2,4\} \\
Kunnen et al. \cite{Kunnen2008, Kunnen2010, Rajaei2016,Joshi2016, Rajaei2017}&[5.68,7.85]&[1.28\times10^8,3.5\times10^9]&[1.17\times10^{-6},8.7\times10^{-4}]&[1.17\times10^{-6},0.56]&[0.023,11.5]&1 \\
Ahlers et al. \cite{Zhong2009, Weiss2010, Weiss2011}&[4.38,6.26]&[2.37\times10^8,7.2\times10^{10}]&[10^{-6},3.36\times10^{-3}]&[3.3\times10^{-7},0.05]&[0.06,22.2]&0.5,1,2 \\
Niemela \& Ecke \cite{Niemela2010,Ecke2014} & 0.7 & [4 \times10^{9}, 4\times10^{11}]& [2\times 10^{-7}, 3 \times 10^{-5}] & [4.6\times10^{-5},2.0\times10^{-3}]& [0.07,5.0] &  0.5  \\
Aurnou et al. (NoMag) \cite{Hawkins2023} & [4,7]& [10^5, 10^{13}]& [10^{-8}, 10^{-2}] & [5.1\times10^{-5},0.20] & [10^{-6},10^3] &  \{0.11, 0.25, 0.33,0.5, 0.75,1.5,2,3,6 \} \\
Bodenschatz et al. (U-Boot) \cite{Zhang2020,Wedi2021} &  0.8 & [8\times 10^9, 8\times 10^{14}] & [1 \times 10^{-8}, 3\times 10^{-5}] & [4.48\times10^{-7},0.19]& [0.07, 42] & \{0.5, 1\} \\
Kunnen et al. (TROCONVEX) \cite{cheng2018_gafd, Cheng2020,  Kunnen2021}  & [2,7] & [1 \times10^{10},7\times 10^{13}] & [5 \times 10^{-9},3\times10^{-6}] & [9.2\times10^{-4},0.20]& [2 \times 10^{-4},30] & \{0.1, 0.13, 0.2, 0.5\}  \\
Abbate \& Aurnou ( Calimero) \cite{Abbate2023} & \{6,41,206,993\} &[10^8,10^{12}] & [10^{-7},10^{-4}] & \{6.9 \times 10^{-5},0.13\}  & [0.02,2.5] & \{0.24, 0.51, 1.0\} \\
Zhong et al. \cite{Ding2021, Lu2021, Shi2020}&4.38&[1.4\times10^7,2.9\times10^{11}]&[2\times10^{-7},1.3\times10^{-4}]&[0.005, 0.31]&[1.2\times10^{-6},6.94]&0.4,0.5,1,2,3.8 \\
Xia et al. \cite{Hu2022}&4.34&[2.8\times10^8,9.5\times10^9]& [1.4\times10^{-6},1.4\times 10^{-5}] &[0.004,0.363]&[0.025,1]&1 \\ \midrule
Chamberlain \& Carrigan \cite{chamberlain1986_gafd} & 6.5 &[3.3\times10^5,2.6\times10^6]&[10^{-6},5.0\times10^{-6}]&[2.0,49]&[2.3\times10^{-4},3.2\times10^{-3}]&1\\
Cordero \& Busse \cite{cordero1992_grl, Cordero1993} & 6.2& [5.3\times10^5,5.1\times10^6]&3.3\times10^{-6},5.6\times10^{-6}&2.1,3.6&9.5\times10^{-4},5.1\times10^{-3}&1.7,3.7\\
Cardin \& Olson \cite{cardin1992_grl,cardin1994_pepi} & 7&4.5\times10^9&[1.2\times10^{-6},1.9\times10^{-6}]&[11,24]&\leq4.8\times10^{-2}&0.5\\
Sumita \& Olson  \cite{sumita1999_sci,sumita2000_pepi,sumita2002_jgr,sumita2003_jfm} & 7&[4.5\times10^8,1.2\times10^9]&2.4\times10^{-6}&7.0&[1.9\times10^{-2},3.8\times10^{-2}]&0.5\\
Aubert et al. (water) \cite{aubert2001_pepi}& 7&[2.9\times10^7,8.6\times10^9]&[1.2\times10^{-6},4.9\times10^{-6}]&[4.9,79]&[0,1.7\times10^{-1}]&0.64\\
Aubert et al. (gallium) \cite{aubert2001_pepi}& 0.025&[1.2\times10^7,2.6\times10^8]&[3.6\times10^{-7},7.2\times10^{-7}]&[20,79]&[0,7.4\times10^{-2}]&0.64\\
Gillet et al. (water) \cite{gillet2007_jfm1}  & 7& [3.3\times10^8,9.1\times10^9] &[1.2\times10^{-6},4.9\times10^{-6}]&[4.9,79]&[8.4\times10^{-3},1.8\times10^{-1}]&0.64\\
Gillet et al. (gallium)  \cite{gillet2007_jfm1, gillet2007_jfm2} & 0.025&[2.4\times10^7,2.3\times10^8]&[4.9\times10^{-7},1.5\times10^{-6}]&[4.9,79]&[1.5\times10^{-2},1.4\times10^{-1}]&0.64\\
Sun et al. \cite{jiang2020_sciadv, jiang2022_prl}&4.34& [10^6,3.7\times10^{12}]&[6.7
\times10^{-7},2.4\times10^{-2}]&[5\times10^{-8},80]&[2.5\times10^{-3},100]&1\\ 
\midrule
Aurnou \& Olson \cite{aurnou2003_epsl} & 7& [3\times10^6,3\times10^{10}]&[10^{-5},9\times10^{-4}]&[1.5\times10^{-5},4.8\times10^{-2}]&[6.6\times10^{-3},59]&[0.66,0.86](0.33)\\
Poth\'erat et al. (LEE1) \cite{aujogue2016_rsi, aujogue2018_jfm, potherat2023_prl}& \{7, 12\} &[1.4\times10^7,2.9\times10^9]&[1.3\times10^{-6},4.5\times10^{-5}]&[2.8\times10^{-2},0.44]& [3.2\times10^{-2},0.58]&0.8(0.35)\\
Poth\'erat et al. (LEE2) \cite{agrawal2024_gji}  & 6.4&[2.1\times10^7,1.1\times10^9]&[3.5\times10^{-6},4.2\times10^{-5}]&0,[3.1\times10^{-3},0.44]&[6.3\times10^{-3},0.29]&1.1\\
\bottomrule\\
\end{tabular}}
}
\end{landscape}

\begin{table}
\caption{Prandtl number $Pr = \nu/\kappa$, Rayleigh number, Ekman number $Ek = \nu/(2\Omega H^2)$, Froude number $Fr = \Omega^2 R_o/g$ (where $R_o = D/2$ in cylindrical geometries), and the (convective) Rossby number $Ro = \sqrt{Ek^2Ra/Pr}$. The aspect ratio $\Gamma = R_i/(R_o - R_i)$ for spheresl and $\Gamma = 2 R_o /H$ for cylinders. Note that $Pr$, $Ra$, $Ek$, $Fr$, $Ro$ are only approximate due to the temperature-dependence of the material properties. For terrestrial gravity we use $g =  \SI{9.8}{m/s^2}$. Most values for cylindrical geometries were taken from the review by Cheng et al. \cite{cheng2018_gafd}. 
No parameters are given in Busse \& Carrigan (1976). 
In the experiments by Boubnov \& Golitsyn  \cite{Boubnov1986, Boubnov1990}, Fernando \cite{Fernando1991} and Sahoo \& Sreenivasan \cite{sahoo2020_epsl, sahoo2020_jfm} convection is driven by imposing  a heat flux instead of a temperature difference.  
Not all values in this table are taken from the original papers, some are estimated and may not represent actual measurement points. 
\label{tab:nondim_parameters}}
\end{table}
\section{Central force field: Outer space and elsewhere \label{sec:central}}
\subsection{Outer space: The Geophysical  Fluid Flow Cell (GFFC) and the GeoFlow experiments}
\begin{figure}
\includegraphics[width=\textwidth]{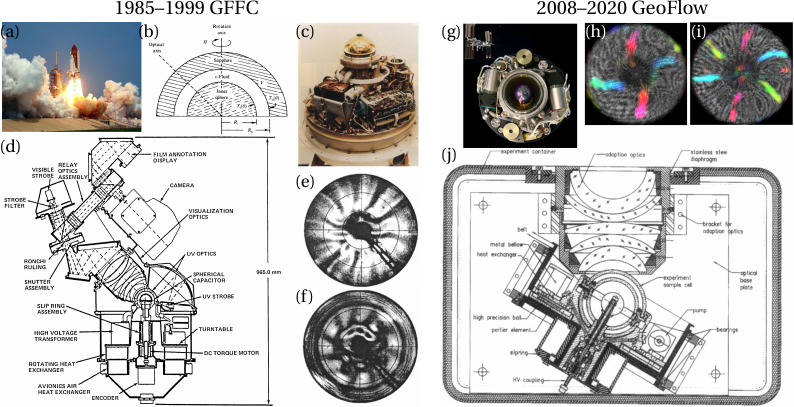}
 \caption{Left panels (a--f): the Geophysical Fluid Flow Cell (GFFC) experiment; (a) Space Shuttle Challenger with Spacelab 3 that flew GFFC to space; (b) sketch of the cross-section of the GFFC test cell \cite{hart1986_jfm}; (c) rotating turntable with power supply, circulation pumps, test cell, and associated circuitry  \cite{hart1986_jfm}; (d) the full GFFC experiment including optical imaging system \cite{Hill1982}; (e,f) unwrapped convection planforms showing banana cells and pronounced polar disturbances, the tangent cylinder is at $43^\circ$, (e) shows north-south fringes (sensitive to latitudinal temperature gradients) and (f) shows east-west fringes (sensitive to longitudinal temperature gradients) \cite{Hart1999}. Right panels (g--j): the GeoFlow experiment; (g) photo of the experimental setup with the International Space Station (ISS), image by ESA; (h,i) Wollaston Shearing Interferometry which visualises the thermal gradients based on the changes of the temperature-dependent refractive index, the columnar cells are coloured using automatic pattern recognition \cite{Zaussinger2020}, (h) spiralling columnar cells and (i) almost straight cells; (j) the full GeoFlow experiment with adaption optics and rotary tray \cite{egbers2003_asr}.\label{fig:space}}
\end{figure}

Spherical geometries with a central force field may at first glance be essential for creating a laboratory model of Earth's outer core, and other geophysical and astrophysical systems. Unfortunately, on the ground, the terrestrial gravitational potential prevents such a setup. Thus, realistic experiments where the angle between the Coriolis and buoyancy force varies, are not straightforwardly realisable. 
We first focus on the two experiments that went to space to escape the terrestrial limitations: the {Geophysical Fluid Flow Cell (GFFC)} \cite{hart1986_sci, Hill1982, hart1986_jfm, hart1996_3mg} and the {GeoFlow} experiment \cite{egbers1999_mst, egbers2003_asr, Futterer2010, Futterer2012, Futterer2013, Zaussinger2017, Zaussinger2020}, shown in figure~\ref{fig:space}. 
Neither experiment specifically aimed at understanding Earth's core, but both constitute unique laboratory experiments of (hemi-)spherical rotating convection. 
%
%
%

The main underlying principle of both the space experiments is that if an electric potential is applied to a dielectric fluid, the neutral non-charged molecules are polarised, and a force acts in the direction of the strongest electric field region \cite{egbers1999_mst, hart1986_jfm}. The idea was originally suggested by Smylie \cite{Smylie1966, Chandra1972} but Hart et al.\cite{hart1986_sci} later independently realised it for the first time in a spherical experiment. These polarisation forces are also known as dielectrophoretic forces and are density- and permittivity-dependent. They act in a very similar manner to traditional buoyancy forces resulting from either the gravitational or centrifugal potential, but here the potential is $\vec{E} \cdot \vec{E}$, with $\vec{E}$ being the electric field. The key advantage is that in spherical geometries these forces act effectively like a central gravity. Hence, in a rotating experiment, the angle between the electro-hydrodynamic ``gravity'' vector and the rotation vector vary latitudinally in a similar manner as in planetary and stellar interiors.

More specifically, one can approximate the functional dependence of the permittivity to the first order, as for the density in the Oberbeck-Boussinesq approximation \eqref{eq:rho_oberbou}:
\begin{equation}
\epsilon = {\epsilon_m}(1 - \gamma (T-T_m)),    
\end{equation}
where ${\epsilon_m}$ is the ambient permittivity and $\gamma$ is the dielectric variability. 
The values for the different fluids used in the space experiment are given in table~\ref{tab:spaceparameters}. Since in all fluids $\gamma \approx \alpha$, the resulting system can be treated with an equivalent version of the Oberbeck--Boussinesq approximation \cite{Oberbeck1879, Boussinesq1903}.
Further, if the electrical field $\vec{E}$ behaves like a spherical capacitor \cite{Futterer2013}, we have
\begin{equation}
\vec{E}(\vec{r}) = \frac{R_i R_o}{R_o - R_i} \frac{1}{r^2} V_{\mathrm{rms}} \hat{\vec{e}}_r,
\end{equation}
where  $V_{\mathrm{rms}}$ is the applied AC voltage to the shells. The resulting electro-hydrodynamic ``gravity'' is then
\begin{equation}
 \vec{F}_{e} = g_{e} \gamma T  \hat{\vec{e}}_r \mbox{ with }  g_{e}(r) = \frac{2 \epsilon_m V_{\mathrm{rms}}^2}{\rho_m} \left(\frac{R_i R_o}{R_o - R_i}\right)^2\frac{1}{r^5}.
\end{equation}
Thus, gravity varies radially but falls off much steeper with $1/r^5$ compared to planetary gravities, closer to {$r$ or} $1/r^2$. This, however, is argued to be only of secondary importance \cite{hart1986_jfm} even though the inner boundary gravity is much stronger than at the outer boundary, i.e. $g_i \gg g_o$. These values are given in table~\ref{tab:spaceparameters}. $Ra$ is the usual Rayleigh number but is defined using the outer value $g_o$. 
The typical values of $g_o$ are much smaller than Earth's gravity. This demands microgravity and severely constrains 
rotation rates and vessel size to small values. 

The \emph{Geophysical  Fluid Flow Cell (GFFC)} flew to space for the first time on the 29\textsuperscript{th} of April 1985 together with two squirrel monkeys, 24 rats and seven astronauts \cite{Hill1982, hart1986_sci, hart1986_jfm}. The principal investigator was John Hart of the University of Colorado at Boulder, it was managed by NASA's Marshall Space Flight Center. The experiment was conducted in the Spacelab, a microgravity laboratory developed by ESA and flown by the NASA space shuttle Challenger. The specific mission was Spacelab 3 (STS 51-B; 29 April--6 May 6, 1985), the first in which a strict low-gravity environment was maintained in orbit, here in a low Earth orbit of \SI{400}{km} altitude. {GFFC}'s second space flight in 1995 was with the US Microgravity Laboratory-2 (USML-2) on board the space shuttle Columbia  (STS-73; 20 October-- 5 November 1995) \cite{Hart1999, hart1996_3mg}. The first mission provided about 110 hours of experimental data mainly in the form of about 50,000 film images. During the second mission, 29 separate 6-hour runs (174 hours in total) were carried out.

The set-up itself is a spherical shell rotating at up to \SI{3}{rad/s}, but only its northern hemisphere was used in the experiment, as depicted in figure~\ref{fig:space}.  
The inner shell was made of polished nickel of radius $R_i = \SI{2.402}{cm}$ 
and the outer northern hemispherical shell of radius $R_o = \SI{3.300}{cm}$ was a $\SI{1}{cm}$ thick transparent sapphire dome, with electrically conducting coating on the inner side acting as an electrode. An AC voltage $V_{\mathrm{rms}} = \SI{10}{kV}\pm 0.5\%$ at $\SI{300}{Hz}$ was applied across the gap.
This ensured the absence of conductive electric currents that may lead to other and undesired fluid instabilities, as the associated period was short compared to the charges' relaxation time of the working fluid. The southern hemisphere was filled with Teflon to avoid the effect of non-radial electric fields occurring in the vicinity of the inner sphere's mechanical support. The working fluid was a dielectric silicone oil Dow Corning 
with $Pr = 8.4$.
The inner sphere was heated and the outer sphere was cooled with a computer-regulated temperature gradient of up to $\SI{20}{K}\pm\SI{0.1}{K}$.
Individual heater elements also allowed to study differential, latitudinally varying heating, and several experiments were run with a hotter north pole (a cooler pole would have obscured visualisations).
During the second flight, problems with the temperature regulation 
led to a temperature drift of about 10\% to 20\% and overheating occurred This limited the data interpretability, especially for the planned different latitudinal heating scenarios. 
As optical techniques shadowgraph and back-focus Schlieren visualisations were used, which give information about the temperature structures. 
Successive images from the fixed camera only allowed flow reconstruction 
in weakly time-dependent flows but not for more chaotic and turbulent flows.
During the second flight, an additional video camera was installed, but it failed at the 19\textsuperscript{th} run. Originally, it was also planned to use spiropyrane, a photochromic dye that is activated by UV light, as a flow tracer for velocity measurements during the Spacelab 3 mission. However, the dye degraded over the time of 1.5 years when it was not accessible during integration into the space module, thus, not permitting the extraction of accurate velocity data.

The {GeoFlow} experiment was very similar in conception, 
and was integrated into the Fluid Science Laboratory of the European Columbus module on 
the International Space Station (ISS). This crucial difference allowed for much longer, extensive campaigns with more than 2,500 hours of scientific run time \cite{egbers1999_mst, egbers2003_asr, Futterer2010, Futterer2012, Futterer2013, Zaussinger2017, Zaussinger2020}. It allowed daily data transfer from the orbit and thereby almost real-time data analysis. The GeoFlow experiment went to space for the first time on 7 August 2008 and stayed there until January 2009  ({GeoFlow I}), with three follow-up missions (March 2011--May 2012; December 2012--May 2013; November 2016--February 2017; ({GeoFlow II, IIb, IIc}). The head of the GeoFlow Team was Christoph Egbers from the Brandenburg University of Technology Cottbus-Senftenberg. Unlike {GFFC}, GeoFlow was a full sphere with inherent limitations due to the heating supply shaft at the south pole \cite{egbers1999_mst, egbers2003_asr}. The device had an inner radius of $R_i = \SI{1.35}{cm}$ and an outer radius of $R_o = \SI{2.70}{cm}$ with the possible rotation rates of $\Omega = \SI{63}{rad/s}$. The working fluid of {GeoFlow I} was the silicone oil M5 with $Pr = 64$. For {GeoFlow II}, the team improved the olfactory aspect and used the alcohol 1-nonanol, known in the cosmetics industry for its lemon fragrance, with $Pr = 125$ and $Pr = 178$. Most of the results were for very slow rotation rates that were mainly there to ensure the visualisation of the full latitudinal direction 
and more relevant to mantle convection \cite{Futterer2010, Futterer2012, Futterer2013, Zaussinger2017}. 
Only the {GeoFlow IIc} experiment focused on rotating convection with parameters comparable to the {GFFC} experiment, with a minimum $Ek = 2.64 \times 10^{-3}$, and maximum $Ra = 1.59 \times 10^5$. The fluid 1-nonanol also allowed them to study both convection induced by the temperature difference between the spherical shells and due to the internal dielectric heating \cite{Zaussinger2020}. The flow visualisation technique used was Wollaston prism shearing interferometry which showed only projections of the thermal structure since safety and weight limitations did not permit the use of tracer particles or larger optical systems \cite{Zaussinger2020}. Exemplar pictures are shown in figure \ref{fig:space}.

\begin{table}[pb!]
\renewcommand{\arraystretch}{1.4}
\setlength{\tabcolsep}{3.1pt}
\resizebox{\textwidth}{!}
{
\def~{\hphantom{-}}
\begin{tabular}{cMMMMMMMMMMMM}
\toprule
fluid & \rho_m [\mathrm{kg/m^3}] & \epsilon_m & \alpha [\mathrm{K^{-1}}] & \gamma [\mathrm{K^{-1}}]  & V_{\mathrm{rms}} [\mathrm{kV}] & g_o [\mathrm{m/s^2}] & g_i [\mathrm{m/s^2}] \\[3pt]
\midrule
 silicone oil DC \SI{0.65}{cs}& 760 & 2.5 \epsilon_0 & 1.34 \times 10^{-3} &  1.29 \times 10^{-3} & 10 &  1.13 & 5.55 \\
 silicone oil M5 & 920 & 2.70\epsilon_0  & 1.08  \times 10^{-3} &  1.07 \times 10^{-3} & 10 & 0.26 & 8.54 \\
 1-nonanol (\SI{20.0}{\degC}) & 820 & 8.60\epsilon_0  &9.45 \times 10^{-4} & 2.95 \times 10^{-3} & 6.5 & 0.40 & 12.76 \\
 1-nonanol (\SI{30.5}{\degC}) & 822 & 8.33\epsilon_0 & 9.45 \times 10^{-4} & 2.46 \times 10^{-3} & 6.5 & 0.39 & 12.32 \\
 \bottomrule\\
\end{tabular}}
\caption{Material properties used in the space experiments, $\rho_m$ is the mean density, $\epsilon_m$ the  mean permittivity,  $\alpha$ the expansion coefficient, $\gamma$ the dielectric variability,$V_{\mathrm{rms}}$ the (maximum) applied a.c. voltage, and $g_o$ and $g_i$ the values of the electro-hydrodynamic ``gravity'' at the outer and inner radius of the (hemi)spherical shell, respectively.
The silicone oil Dow Corning \SI{0.65}{cSt} 200 Fluid was used in the {GFFC} experiment, silicone oil M5 in the {GeoFlow I} experiments, and 1-nonanol at different mean temperatures during the {GeoFlow II} experiments. Note that $\epsilon_0 = 8.854 \times \SI{10^{-12}}{C \, V^{-1} m^{-1}}$ is the vacuum permittivity.
Taken from references \cite{hart1986_jfm} and \cite{Futterer2013, Zaussinger2017}. \label{tab:spaceparameters}}
\end{table}

Both experiments, {GFFC} and GeoFlow studied the stability properties, the formation of patterns and the transition between states, including chaos and turbulence.
Many of the results had rather high values of $Ek$ and $Ro$, as well as low values of $Ra$, 
thus, they are 
practically in a rotation-unaffected regime \cite{Aurnou2020, kunnen2021_jot, Ecke2023}. These flows show a tessellated pattern, also found in more recent numerical simulations of non-rotating convection \cite{Gastine2015, Featherstone2016}.
The \nohyphens{GeoFlow} experiments also show 
mantle-like plumes influenced by non-Oberbeck-Boussinesq effects \cite{Horn2013, Futterer2010, Futterer2012, Futterer2013}.

In the rotating regime, the main result was the confirmation of the existence of columnar convection in radial gravity (``banana cells'' \cite{busse1970_jfm}) in the equatorial region. Experiments also captured their drifting in the prograde direction, illustrated in figure~\ref{fig:space} \cite{hart1986_jfm, hart1986_sci, Zaussinger2020}. In both experiments, columnar convection was very robust, likely because of the relatively high Prandtl number \cite{Julien2012, Stevens2010b, Horn2014, Horn2015, kunnen2021_jot, Abbate2023}. The experiments, especially {GFFC}, also showed that at higher supercriticalities, retrograde mid-latitude convection modes develop, the TC becomes convectively active, and later on these mid-latitude and polar modes become unstable.
Likely, there were two effects at play here, a loosening of the rotational constraint, i.e. moving closer to the non-rotating regime expected at higher supercriticality \cite{agrawal2024_gji}.
The unique specificity of {GFFC} was that they could also observe the interactions of the columnar equatorial cells with the retrograde-propagating mid-latitude convection modes, which can only be captured with spherical radial gravity. These interactions initially caused a wavering, and finally the erosion of the columnar equatorial cells. They also found that the virtual boundary of the tangent cylinder broke down, with columnar convection only within 20$^\circ$ of the equator whereas the tangent cylinder extended up to 43$^\circ$. The no-slip boundary conditions (in addition to the not-very-extreme control parameters) were likely the reason for not observing any zonal flows. However, these boundary conditions are arguably also more relevant for Earth's core than for gas giants. Many of the results are indeed reminiscent of more modern numerical spherical simulations of rotating thermal convection \cite{Gastine2016, Featherstone2016, Featherstone2023}. They further also showed that spherical shell thermal convection may be hysteretic, finding sometimes the same and sometimes different flow patterns depending on whether they increased or decreased $Ra$ \cite{hart1996_3mg, Hart1999}.
The {GFFC} experiment also explored latitudinal inhomogeneous heating. They observed spiralling convection when the north pole was heated, which was interpreted as evidence for baroclinic waves. They showed that with increased differential heating, the interaction of the equatorial columnar cells with mid-latitude waves leads to triangular waves and ultimately, with turbulent structures moving downward from the pole to full turbulence.

\subsection{... and elsewhere: Kelvin forces and pycnoclinic acoustic gravity}
The main limitations of the space experiments' applicability to planetary core 
convection are the relatively low $Ra$ and high $Ek$ required to minimise centrifugal effects. 
Their other shortcoming 
is that the poor electric conductivity of dielectric materials 
renders the study of MHD effects practically infeasible.

Having said that, it may be worthwhile to embrace Coriolis-centrifugal convection, and so thanks to the combination of $\Omega^2\vec{r}$ and $\vec{g}$, spot tornado-like vortices as they occur in planetary and stellar atmospheres \cite{Luna2015, Horn2018, horn2019_prf, Horn2021, Ding2021, Kuniyoshi2023, Tziotziou2023}.

Ferrofluids offer an alternative route to creating a central force field, with the benefit that some of them are also electrically conducting.
Experiments of this kind have been conducted in non-rotating thermo-magnetic convection where spherical gravity is mimicked using permanent magnets as an inner core, and a ferrofluid fills the outer core \cite{Rosensweig1999, frueh2005_nlpg}. 
However, neither rotating convection nor magnetoconvection have so far been realised.
The idea is here that a magnetised fluid is attracted towards higher magnetic fields. This attracting body force is referred to as Kelvin force and mimics buoyancy. The magnetisation depends on both the magnetic field and the temperature, a colder fluid is attracted more strongly than a warmer fluid.
This is expressed through the pyromagnetic coefficient which plays the role of  
the expansion coefficient.

Acoustic gravity is another clever way of creating thermal rotating convection with a central force field in an Earth-bound laboratory setting \cite{Koulakis2023}:
Sound, when averaged over many cycles, exerts a force on density gradients in a gas.
The so-called pycnoclinic acoustic radiation force $-\langle v_{ac}^2\rangle \vec{\nabla} \bar{\rho}/2$ yields an acoustic gravity of $\vec{g}_{ac} = \vec{\nabla}\langle v_{ac}^2\rangle/2$ \cite{Koulakis2018, Koulakis2021}. Here, $\bar{\rho}$ is the time-averaged density and $v_{ac}$ is the acoustic velocity. 
Koulakis et al. \cite{Koulakis2018} use a rotating spherical plasma bulb with a radius of $\SI{1.5}{cm}$ filled with a weakly ionized sulfur gas. 
The gas is heated volumetrically by microwaves. The rotation is essential to guarantee quiescent conditions of the plasma. Then, the amplitude modulation of the microwave power generates a high amplitude, spherically symmetric acoustic standing wave. This creates an acoustic gravity that can briefly reach values of up to 1000 times Earth's gravity, so centrifugal effects become negligible. 
Gravity changes sign roughly at half the bulb radius, so confining the convective zone to the outer region, as in stars and Earth's core. 
While the parameters are only slightly more extreme than the space experiments,
 density differences of a factor of about two make it possible to study compressible rotating spherical convection too. 

\section{Cylindrical radial gravity \label{sec:cyl}}
Given the difficulties in modelling spherical rotating thermal convection with central force fields, geophysicists, planetary scientists and fluid dynamicists had to resort to alternative means to model planetary core convection. A very successful branch of laboratory research in this regard relied on centrifugal buoyancy 
to mimic equatorial and low-latitude convection.

\subsection{The Busse Annulus}
The cylindrical annulus geometry was first conceptualised as a laboratory device with direct application to Earth's core by Busse in 1970 \cite{busse1970_jfm}. The underlying idea here is to use the cylindrically radially outward-directed centrifugal force and then cool the inner boundary and heat the outer one. This ensures that buoyancy acts in the right direction, as sketched in figure~\ref{fig:geometries}. Only the product of gravity and temperature gradient matters physically. 
Busse let the centrifugal acceleration exceed Earth's gravity by a factor of two or three \cite{Busse2009} and envisioned slanted boundaries, where the height varies with cylindrical radial distance from the vertical rotation axis. The resulting $\beta$-effect creates slowly drifting columnar thermal Rossby waves. These are nowadays also known as \emph{Busse} columns and have been argued to exist in planetary cores \cite{busse1970_jfm}. Busse first derived a simple analytical model for the onset Rayleigh number, drift frequency and wave number in the small gap approximation with asymptotically small endwall slopes.
He then materialised his theoretical ideas into an innovative laboratory setup \cite{busse1974_jfm}.  
The concept of the \emph{Busse annulus} turns the arduous three-dimensional spherical problem into a tractable two-dimensional one and offers simple access to the spherical rotating convection problem.

Busse (and collaborators) subsequently fleshed out the original idea \cite{busse1974_jfm, Azouni1986, Busse1982, Schnaubelt1992, Busse1986a, Busse1986b, Or1987, Westerburg2003, Jaletzky2000, Tao2006, Auer1995}, including the application to magnetoconvection \cite{Busse1993, Kurt2007} and the geodynamo \cite{Busse1975} (see also the reviews \cite{Busse1994, Busse1997, Busse2002, Busse2009} and subsection~\ref{sec:sphere} on the spherical geometry). The \emph{Busse Annulus} has been subject to many theoretical and numerical studies up to recent days \cite{Bassom1996, Oishi2022, Teed2012, Wicht2018, Calkins2013, Barik2023}. In particular, Calkins et al. \cite{Calkins2013} lifted the original restriction to small slopes and near-criticality. He introduced an asymptotically reduced three-dimensional set of equations for quasi-geostrophic convection, thus, establishing a closer connection to spheres and spherical shells. 

\begin{figure}
\includegraphics[width=\textwidth]{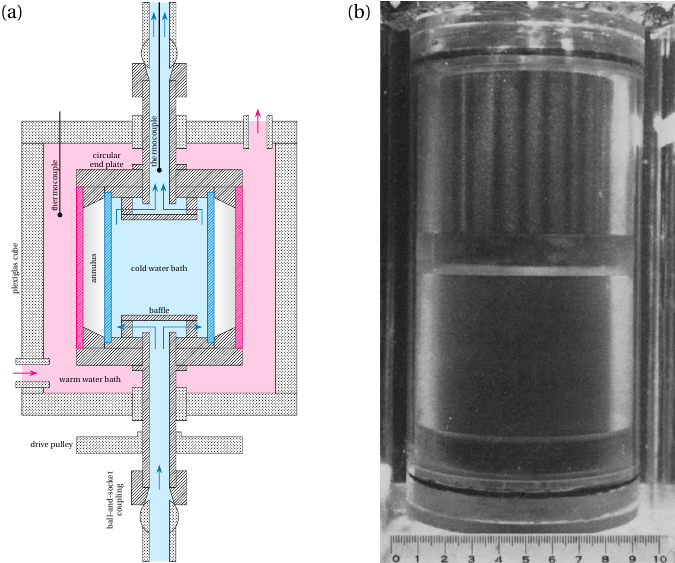}
 \caption{(a) Schematic vertical cross-section of Busse and Carrigan's original annulus apparatus, adapted from ref. \cite{busse1974_jfm}. The northeast lines pattern indicates the rotating parts and the crosshatch dotted pattern indicates the non-rotating parts, in particular, the Plexiglas cube. The radius and height of the inner and outer cylinders were adaptable and varied during the experimental campaigns. (b) Photograph of two annuli stacked on top of each other, scale is in cm, adapted from ref. \cite{busse1974_jfm}. The upper part shows an annulus with constant height, showing strong convection columns. The lower part shows an annulus with conical endwalls, that results in an axisymmetric flow at the same parameters. The working fluid was water at \SI{30}{\degC} and Kalliroscope was used for visualisation.\label{fig:busseannulus}}
\end{figure}

Here, we focus on the experimental aspects of the \emph{Busse Annulus} \cite{busse1974_jfm, Busse1982, Azouni1986, Busse1997, Jaletzky2000}.
A sketch of the first laboratory realisation \cite{busse1974_jfm} is shown in figure~\ref{fig:busseannulus} (a). Two concentric cylinders were mounted on two circular end plates. The outer cylinder was heated through a circulating thermostatically controlled reservoir enclosed in an outer nearly cubical Plexiglas box.
The inner cylinder was cooled by circulating water through two centred hollow shafts mounted on the end plates, and ball-and-socket couplings connected the stainless steel shaft to the cooling tubes. Baffles at the inside of the end plates ensured an efficient distribution towards the inner annulus wall. The shafts also served as rotation axis, and rotation rates of up to \SI{400}{rpm} (\SI{41.9}{rad/s}) were achieved. The temperature difference was measured using a thermocouple in each of the water baths. Since the experiment concerned the onset of convection, typical temperature differences were less than \SI{1}{K}.
The inner cylinder was made of aluminium and three different sizes were used, $R_i = \{\SI{13.34}{cm}, \SI{3.81}{cm}, \SI{3.18}{cm}\}$. The outer cylinder was made of acrylic plastic (Plexiglas) to allow for optical access, and the outer Plexiglas cube helped to avoid optical distortions. The used radii were $R_o = \{ \SI{13.94}{cm}, \SI{4.775}{cm}, \SI{4.74}{cm}$\}. 
The endwalls were made of Teflon and three different sloping angles were used: $\chi = 0$ (corresponding to horizontal plates) and $\chi = 22.5^\circ$ and $45^\circ$. The mean heights were $L = \{\SI{28.6}{cm}, \SI{6.97}{cm}, \SI{6.69}{cm}, \SI{1.96}{cm}, \SI{0.95}{cm},\SI{0.45}{cm}\}$. The working fluids were silicon-oil 
and water. Small amounts of Kalliroscope were used for flow visualisation. These are nearly neutrally buoyant particles that align with the shear{. S}troboscopic light was used to make the shear patterns visible and to measure the rotation rate.
The onset of convection was determined visually. 

Quantitative measurements involved measuring a buoyancy parameter $B$ versus an inverse Ekman number $Ek = \nu/(\Omega L^2)$ (bar the factor 2 in both cases). In modern terminology, their buoyancy parameter is in fact the square of the (convective) Rossby number. This can be seen by recalling that $Ro= \tau_{rot}/\tau_{buoy}$, i.e. the ratio of a rotational time scale, $\tau_{rot} = 1/\Omega$ and the buoyancy time scale, $\tau_{buoy} = H/\sqrt{\alpha \Delta \Omega^2 \bar{R} H}$, defined in analogy to the usual free-fall time scale, with the temperature difference $\Delta =T_h - T_c$, $\Omega^2 \bar{R}$ being used as centrifugal gravity, and the characteristic length scale being the gap width $H$. Thus, we have
\begin{equation}
Ro^2 = B= \frac{\tau_{rot}^2}{\tau_{buoy}^2} = \frac{\alpha \Delta \Omega^2 \bar{R} H}{H^2 \Omega^2} = \frac{\alpha \Delta \bar{R}}{H}.
\end{equation}
At the onset of convection, Busse \& Carrigan saw the first experimental evidence of the retrograde drifting columns that were slightly tilted and helical.
Their results showed decent agreement with the small gap linear theory \cite{busse1970_jfm}. Qualitatively, it was also found that the wavelength of the instability decreased with increasing rotation rate in the sloping endwall cases. The instabilities in this case were typically the superposition of several waves and the columns showed beating phenomena.
%
%
%
%
Dedicated experiments sought hysteresis and subcritical convection but found none of them.
The $\beta$-effect arising out of the interplay of the Coriolis force with the top and bottom conical endwalls was found to inhibit the convective instability  
compared to the flat case that develops columns, see figure~\ref{fig:busseannulus} (b).
At high rotation rates the experiment strikingly exhibits geostrophic columns, in agreement with the TPC, which imposes that the flow be $z-$independent in these regimes.
\paragraph{Zonal flows}
For the observations of zonal flows, Busse \& Hood modified the Busse annulus 
($L = \SI{12.6}{cm}$, $H = \SI{2.46}{cm}$, $\bar{R} = \SI{3.51}{cm}$) to include radial curvature of top and bottom boundaries, either convex or concave \cite{Busse1982}. 
The mean flow was measured by releasing electrolytic dye into the working fluid (water) by frequent pulsed currents, following a method devised by Baker \cite{Baker1966}, and taking photographs at intervals of \SI{10}{s}.

A similar experiment by Azouni et al. \cite{Azouni1986} used mercury and water as as working fluids. The outer cylinder was replaced by anodized aluminium to avoid the thermal loss through the acrylic sidewall 
($L = \SI{14.48}{cm}$, $\bar{R} = \SI{3.917}{cm}$, $H = \SI{3.917}{cm}$ and $\chi = 47.4^\circ$). They measured drift rates and amplitudes of the convection columns by combining and correlating five thermistor probes in the equatorial plane and spaced in azimuth. Three were embedded in the inner cylinder flush with the outer surface, two others were glued to the outer cylinder. 

Earth's gravity creates a thermal wind and a meridional circulation scaling as $\Omega^{-1}$ and $\Omega^{-1/2}$, respectively.
The thermal wind is antisymmetric relative to the equatorial plane, being retrograde in the lower part and prograde in the upper part, thus, it does not affect the dye injected in the equatorial plane too much.
The meridional circulation is noticeable, on the other hand. It leads to an upflow near the hot outer wall and a downflow near the cold inner wall, thus, there is an additional prograde zonal flow at the outer wall and a retrograde flow at the inner one. 

\begin{figure}
\includegraphics[width=\textwidth]{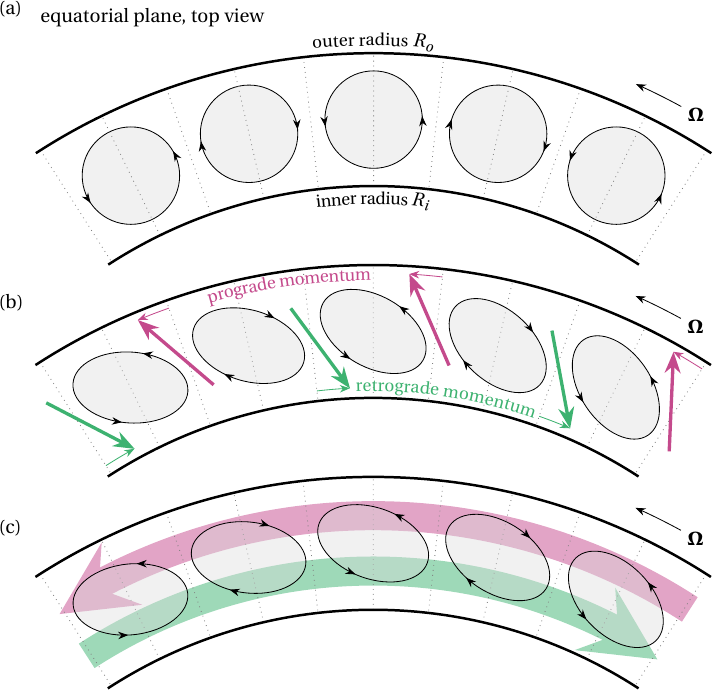}
 \caption{Creation of mean zonal flows in spherical shell and annuli geometries. (a) Convective columns, initially at rest; (b) Columns get tilted, here in the prograde direction, e.g. by small fluctuations which create Reynolds stresses. Prograde momentum (purple arrows) is carried outwards and retrograde momentum (green arrows) is transported inwards. (c) Resulting prograde flow at the outer boundary and retrograde flows at the inner boundary increases the initial tilt leading to a feedback process sustaining a mean flow.  \label{fig:diffrot}}
\end{figure}

Both experiments confirmed the existence of a mean zonal flow. 
The main differential rotation was argued to stem from a mean flow instability due to Reynolds stresses generated through strong mean zonal shear, shown schematically in figure~\ref{fig:diffrot}. The convective columns in figure~\ref{fig:diffrot}(a) get slightly tilted, e.g. by small fluctuations. This creates Reynolds stresses $\langle u_r' u_\phi' \rangle$, where $u_r'$ and $u_\phi'$ are the fluctuating part of the radial and azimuthal velocity component, respectively. Depending on the initial tilt, either direction of the evolving mean flow is possible. If tilted in the prograde direction, figure~\ref{fig:diffrot}(b), then prograde momentum (purple arrows) is carried outwards and retrograde momentum (green arrows) is transported inwards. Thus, the flow is retrograde at the inner boundary and prograde at the outer boundary. The resulting flow increases the initial tilt leading to a feedback process sustaining a mean flow, figure~\ref{fig:diffrot}(c). Thus, there is a differential rotation in which the outer fluid rotates faster than the inner one. If the columns are tilted in the opposite direction, the mean flow direction is reversed. 
For similar reasons, for convex endwalls, thermal Rossby waves propagate faster on the outside than on the inside, thus, the columns spiral outwards and the mean flow is retrograde at the inner boundary and prograde on the outer boundary. The opposite happens for concave endwalls \cite{Busse1982,Busse2009}. 
Viscous stresses oppose the differential rotation and lead to an equilibrium. The differential rotation was argued to be rather insensitive to the cylindrical annulus sidewall, and, hence should also be found in spherical shells \cite{Busse2009}.
Azouni et al. \cite{Azouni1986} found drift rates in agreement with Busse's simplified theory, whereas the amplitudes decreased with increasing Rayleigh number.

\paragraph{More recent developments}
Busse continued investigating flows in the \emph{Busse annulus}.
He and his collaborators found quasi-geostrophic chaotic states in air \cite{Busse1997}.
They also devised an elegant method to visualise convective patterns using thermochromatic liquid crystals in high-$Pr$ fluids \cite{Jaletzky2000}. 
At such high $Pr$, the effects of rotation were much lower. They found different types of patterns, such as knots and hexaroll, which are predicted theoretically \cite{Auer1995}, but also patterns that still lack a theoretical description such as oblique rolls.

The \emph{Busse annulus} without sloping endwalls was later replicated by Jiang et al. \cite{jiang2020_sciadv, jiang2022_prl}; they rediscovered the centrifugal force as a proxy for gravity and called it supergravity. Their achievable parameters were comparable to Busse's original design, with rotation rates between \SI{211}{rpm} to \SI{705}{rpm}, focussing on the strongly supercritical regime. While they also rediscovered the mean flow instability, the constant radial height does not permit e.g. Rossby waves and thus, makes the experiment less geophysically relevant.

\subsection{Spheres, spherical annuli and hemispherical shells \label{sec:sphere}}
Spherical geometries are a natural improvement of the \emph{Busse annuli}, as they capture a more realistic $\beta-$effect and, in some setups, the Stewartson layers wrapped around the TC \cite{greenspan1969,kunnen2013_jfm}.
Here too, centrifugal gravity is used to mimic the gravitational field in the equatorial region, still at the expense of the polar regions.
As for the \emph{Busse annulus}, since the centrifugal gravity points outwards, the temperature difference has to be reversed with the outer spherical wall held at constant hot temperature using either water \cite{carrigan1983_jfm} or air \cite{cardin1994_pepi}. 

In spherical shell geometries, the radial temperature gradient is controlled by cold fluid (water, or even Kerosene \cite{shew2005_pepi}) circulating through the inner sphere. A solid shaft must be fitted to connect the inner sphere to the cooling circuit and to hold the solid sphere at the centre of the outer sphere. This, however, finalises the sacrifice of the polar regions. Because of this shaft, we refer to setups of this category as the \emph{spherical annulus} configuration.
Another non-planetary feature of the spherical shells is the misalignment of the temperature gradient and gravity that increases with latitude. This source of baroclinicity combined with the Coriolis force incurs an artificial azimuthal wind, and results must therefore be interpreted keeping its influence in mind \cite{carrigan1983_jfm}.

\begin{figure}[h!]
\includegraphics[width=\textwidth]{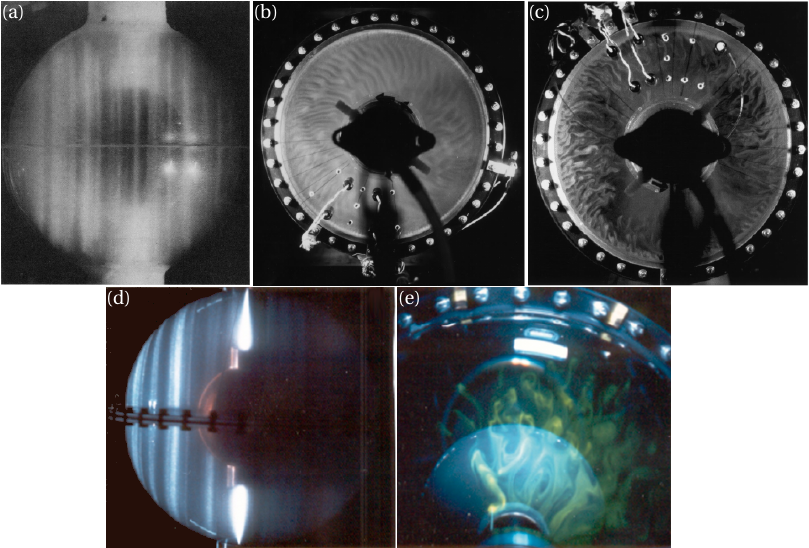}
\caption{Patterns of rotating convection in a spherical gap by order of criticality: (a) Rossby waves at the onset of convection originally observed by Busse \& Carrigan (side view)\cite{busse1976_sci}, (b) spiralling columns at $\widetilde Ra=5.9$ (top view through the transparent top lid of the hemispherical vessel)\cite{sumita2000_pepi} (c) dual convection at $\widetilde Ra=19.3$ (top view as for (b))\cite{sumita2000_pepi} (d) chaotic convection at $\widetilde Ra=50$ (side view) \cite{cardin1994_pepi}, (e) same as (d), but top view.\label{fig:sphere_patterns}}
\end{figure}

Several experiments were built on this principle with variants trying to address some of its shortcomings.
These experiments are classics of geophysical rotating convection and have been extensively reviewed \cite{olson2011_pepi,olson2013_areps,cardin2015_tg}. We shall only recall their main outcomes and highlight some specificities of the experimental approach itself.
\begin{figure}[h!]
\includegraphics[width=\textwidth]{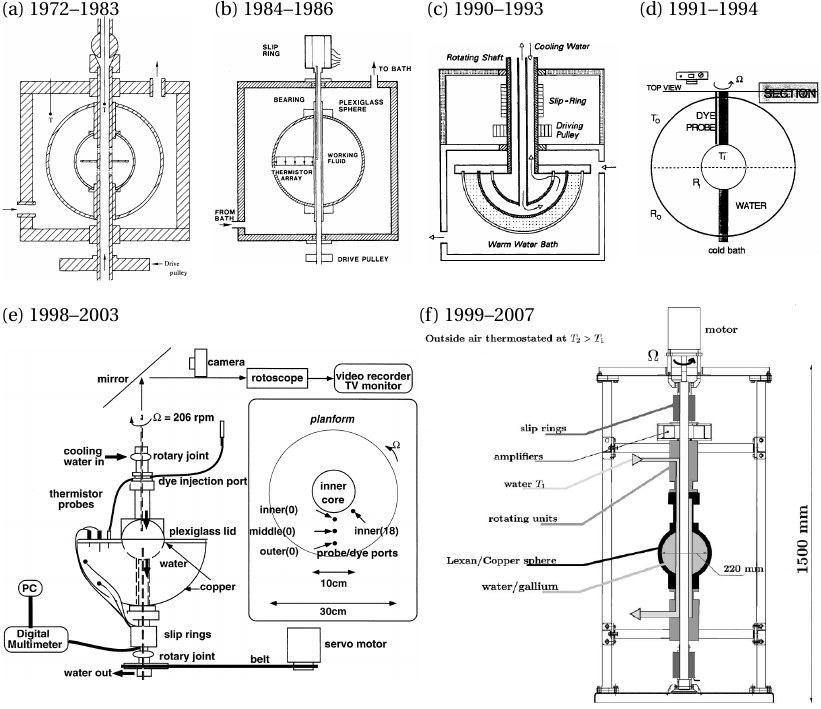}
\caption{Evolution of the spherical shell experiments since 1976:
(a) 1972--1983: Busse \& Carrigan's original spherical shell convection device with spherical inner cold boundary \cite{busse1976_sci,carrigan1983_jfm},
(b) 1984--1986: Chamberlain \& Carrigan's modification of Carrigan \& Busse's original setup, with no inner sphere and linearly increasing outer wall temperature \cite{chamberlain1986_gafd}
(c) 1990--1993: Cordero \& Busse's hemispherical experiment\cite{cordero1992_grl}
(d) 1001--1994: Cardin \& Olson (1994)'s air-cooled setup \cite{cardin1994_pepi}
(e) 1999--2003: Sumita \& Olson's modified version of Cardin \& Olson's setup with a hemispherical annulus \cite{sumita1999_sci,sumita2000_pepi,sumita2002_jgr,sumita2003_jfm}
(f) 1999--2007: Aubert and collaborators' water and liquid metal experiment \cite{aubert2001_pepi,gillet2007_jfm1} \label{fig:sphere_setups}
}
\end{figure}
\paragraph{Experiments at moderate $\boldsymbol{Pr}$: water and oil}
The first spherical annulus experiment was designed by 
Busse \& Carrigan \cite{busse1976_sci,carrigan1983_jfm}, whose original sketch is reproduced in figure \ref{fig:sphere_setups}(a). The working fluids were water and ethylene glycol and the inner core radius was varied by using different inner spheres.
High rotations up to \SI{1000}{rpm} ensured the gravity was effectively cylindrical radial.
The outer sphere (inner radius \SI{10}{cm}) was made of transparent plastic and heated by a temperature-controlled water bath.
The flow was diagnosed by seeding the fluid with reflective platelets that align with the shear and taking snapshots through the outer sphere. From these visualisations, the authors were able to detect the onset of convection and characterise the size of the onset structures.
They mostly recovered Busse's linear theory for the onset of convection in a rotating shell with central gravity in the small-gap limit and identified the onset structures as being Rossby waves originating near the inner sphere, at $Ra_c\sim Ek^{-4/3}$, see original visualisations in figure \ref{fig:sphere_patterns} (a).
The critical wavenumber of the Rossby waves follows the theoretical scaling law but with a lower prefactor.
A discrepancy between the experiment and the small gap theory arises due to the presence of Stewartson layers developing along the TC.
The Stewartson layers tend to stabilise the Rossby waves because they are precisely located where the waves originate. This effect becomes negligible in the limit $Ek\rightarrow0$.

Chamberlain \& Carrigan \cite{chamberlain1986_gafd} later modified the setup, by removing the inner sphere and thereby the Stewartson layers \cite{chamberlain1986_gafd}, see figure \ref{fig:sphere_setups}(b). The outer bath temperature was controlled to increase linearly in time to maintain a constant temperature gradient in the absence of inner cooling. After a short transient, the lag in the temperature response of the fluid inside the sphere leads to a constant temperature gradient there. The radial temperature profile is then the same as if the heat source was distributed in volume, up to a background temperature linearly increasing in time. This radial temperature profile was monitored by radially aligned temperature probes. The critical Rayleigh numbers were found in better agreement with Robert's linear theory for an internally heated sphere \cite{roberts1968_prsa} than Busse's small-gap linear theory \cite{busse1970_jfm}. As in the previous experiment, however, the centrifugal wind incurred by the high-latitude baroclinicity prevented the authors from observing the retrograde drift of the Rossby waves. This led them to suggest experiments using a cylindrical annulus to eliminate that effect as Busse and collaborators did \cite{Busse1982, Azouni1986}.

The first measurement of the drift of Rossby waves with a spherical geometry came from Cordero \& Busse \cite{cordero1992_grl} using a hemispherical shell. Baroclinic effects were minimised by rotating the hemisphere at a speed where the sum of gravity and centrifugal forces produced an apparent gravity with parabolic iso-values nearly parallel to the curvature of the sphere. The outer sphere was made of glass and heated by an external thermostatically-controlled bath of water. The inner sphere was made of brass with an internal water-cooling circuit. 
Using the same visualisation method as in spherical experiments \cite{busse1976_sci} and thermistors, the authors captured the variations of the decay of the drift velocity with the Rayleigh number as well as the appearance of secondary rolls of larger wavelength near the outer sphere.

The next series of spherical annulus experiments was initiated by Cardin \& Olson \cite{cardin1992_grl, cardin1994_pepi}, see figure \ref{fig:sphere_setups}(c). They also used water but with a slightly larger outer sphere of radius \SI{15}{cm}. Unlike previous spherical annuli, the outer temperature was thermostatically controlled at room temperature by the air current induced by the sphere's rotation.
The rotation up to \SI{400}{rpm} was somewhat lower than in Carrigan \& Busse's experiment, so the proxy-gravity was not entirely cylindrical-radial but also had a significant vertical component.
A new visualisation system relied on snapshots synchronised with a strobe light illuminating vertical planes and a fluid seeded with rheoscopic flakes to identify the structures. In the horizontal planes, the flow was visualised by releasing dye at the inner sphere. 
The new visualisation system made it possible to investigate the regimes beyond the onset of convection. They found that for moderate and higher supercriticality, the classical Rossby waves give way to a chaotic columnar regime. The flow is quasi-geostrophic, made of irregularly distributed columns extending over the entire shell height. Its basic pattern is made of retrograde vortices produced at the inner core boundary extending outward into prograde spiralling streets of retrograde vortices (see figure \ref{fig:sphere_patterns} (d)).
This regime is consistent with equatorial mirror-symmetry \cite{hulot1990_jgg} and non-periodic magnetic flux \cite{gubbins1987_nat} inferred from geomagnetic data.
Scaling up experimental velocities in this regime to Earth parameters yields estimates of \SI{0.1}{cm/s} for the convective velocity, consistent with estimates from geomagnetic variations \cite{bloxham1989_ptrsa}. However, the extrapolated column size of \SI{20}{km}-radius columns is unobservable and inconsistent with the large-scale geomagnetic 
data. Cardin \& Olson attributed this discrepancy to the action of the Lorentz force in the core.

The transition to the chaotic regime and the fully developed stages of convection were
further investigated by Sumita \& Olson \cite{sumita2000_pepi}. They modified Cardin \&
Olson's setup into a lower hemisphere, with copper inner and outer boundaries, closed at the
equatorial plane with a transparent Plexiglas lid. The copper parts enabled a better control of the
temperature with a more homogenous distribution at the boundary. The lid offered a
better visualisation window
and its near-thermally insulating properties left the mostly radial conducting thermal gradient unaffected. The temperature was monitored by thermistors inserted at adjustable depths through the lid. They rotated the experiment at a fixed speed of 206 rpm to generate a parabolic gravity iso-potential, with radial gravity seven times greater at the equator than at the pole.
Under these conditions, Sumita \& Olson observed the progressive filling of the gap by Rossby waves drifting in the retrograde direction for $\widetilde{Ra}=Ra/Ra_c\leq 8$, see figure \ref{fig:sphere_patterns}(b), as well as a radial increase in their azimuthal wave number: the effect is driven by the increasing slope enforcing an ever stronger TPC suppressing radial motion.
At higher forcing $\widetilde{Ra} > 8$, this effect splits the flow into a turbulent inner region and strongly suppressed motion in the outer region where Rossby waves are expelled, a regime they coined \emph{dual convection}, illustrated in figure \ref{fig:sphere_patterns} (c). 

Further experiments in the same setup using silicon oil and water as working fluid enabled Sumita \& Olson to explore much higher levels of supercriticality, up to $\widetilde{Ra}=612$ \cite{sumita2003_jfm}.
In these regimes, the turbulent convection homogenises the temperatures to a nearly isothermal state with a thin boundary layer near the inner boundary and the heat transfer follows a scaling law of the form $Nu\sim Ra^{0.4}$. This law, and a comparison of experimental thermal fluctuations to theory \cite{cardin1994_pepi} led the authors to identify this regime as geostrophically turbulent.
\paragraph{Experiments at low $\boldsymbol{Pr}$: liquid metals}
The advent of Ultrasound Doppler Velocimetry  (UDV) \cite{brito2001_ef,baker2017_ef, Vogt2021}, which unlike the name suggests does not rely on the Doppler effect \cite{signalprocessing}, made it possible to measure velocities in the bulk flow of core fluid-like liquid metals.

Aubert and collaborators \cite{aubert2001_pepi} took advantage of this technology to perform experiments in a spherical annulus geometry with a thick central shaft, and no inner sphere, see figure \ref{fig:sphere_setups}(f). The shaft provided cooling through a central cylindrical copper wall, so in planetary terms, Aubert et al.'s experiment had a cylindrical core and a solid, impermeable TC. They performed experiments both in water and gallium.
With water, direct visualisations with Kalliroscope were obtained through the outer sphere made of transparent plastic and maintained at thermostated room temperature by air currents as in Cardin \& Olson's experiment \cite{cardin1994_pepi}. With gallium, optical visualisation is not possible and the outer sphere was replaced by a copper sphere heated by a resisting wire wrapped around it.
They measured velocities in both fluids across the gap with two probes along two directions within the equatorial plane.
From these two measurements, they are able to reconstruct radial (convective) and azimuthal (zonal) velocities.
For both fluids, the maximum convective velocity in the turbulent regime scales as $u_r\sim (\alpha g Q/\rho C_p\Omega^3(R_o-R_i)^4)^{2/5}$, a law that can be recovered by assuming a triple balance between Coriolis, inertia and the Archimedean (buoyancy) force, called CIA balance. Here, $Q$ is the heat flux, and $C_p$ is the heat capacity at constant pressure of the fluid. Velocities rescaled by this law are higher in gallium than water for the same criticality. The reason is the two-order-of-magnitude lower $Pr$ of gallium. 
These high velocities drive a strong zonal flow, 
which forms as the condensate of an inverse energy cascade fed by the convective scales \cite{gillet2007_jfm1}.  The condensate materialises at Rhines' scale, which results from the balance between the $\beta-$effect and Reynolds stresses \cite{rhines1975_jfm}. Energy dissipation at this scale and for the larger scales of turbulence occurs by Ekman friction. However, bold extrapolation of this phenomenology to Earth leads to overestimated velocities by an order of magnitude and too small scales, compared to geomagnetic data. The authors attributed the mismatch, again, to the absence of a magnetic field in their experiments.

Gillet and collaborators conducted further experiments with the same setup up to
$\widetilde Ra=80$ focusing on the zonal flow \cite{gillet2007_jfm1}. 
In the viscous regime of water experiments, the zonal flow scales with the onset convective scales. In the inertial regime reached with gallium, by contrast, they recovered the phenomenology of Aubert et al. \cite{aubert2001_pepi} with a zonal flow scaling as Rhines' scale.
Based on this phenomenology, they showed that at these larger
criticalities, the average zonal flow $\overline U_\phi$ scales with the \emph{rms} radial convective flow $\widetilde U_r$ as $\overline U_\phi\sim \widetilde U_r^{4/3}$. Unlike early experiments at low criticality in water, where the zonal flow expected from the drift of Rossby waves is weak, the zonal flow observed in this regime is at least an order of magnitude faster than the parasitic thermal wind incurred by the misalignment of temperature gradients and gravity.
Hence, in these turbulent regimes at low $Pr$, the spherical annulus produces a more geophysically relevant zonal flow than most of the water experiments.

{Zonal flows controlled by the Rhines scales were also recovered in the \emph{Coraboloid} experiment recently conducted at UCLA \cite{lonner2022_jgrp}: the $\beta-$effect was obtained by rotating water in a \SI{37.25}{cm} diameter cylindrical tank with a free surface. Rotations speeds of \SI{35}{rpm}, \SI{50}{rpm} and \SI{60}{rpm} lead to a parabolic deformation of the free surface. Convection in this 'paraboloid core' was driven by a cold \SI{10.2}{cm} cylinder placed at the centre of the tank spanning the entire fluid height. Zonal flows form in all cases but are more stable at \SI{60}{rpm}. The Rhines scale varies little between \SI{50}{rpm} and \SI{60}{rpm} but centrifugal buoyancy only dominates vertical buoyancy at \SI{60}{rpm}. The authors concluded that centrifugal buoyancy stabilised the zonal flow, which is consistent with Gillet et al's \cite{gillet2007_jfm1} experiments in liquid metal. The properties of jets in a geophysical context are extensively reviewed by Read et al. \cite{read2024_crphys}.} 

Lastly, Shew \& Lathrop \cite{shew2005_pepi} built a large spherical annulus (external diameter $D=\SI{60}{cm}$), with magnetic runs and dynamo applications in mind. 
A much larger setup was later built more specifically to create dynamo, which is still under development \cite{lathrop2011_pd}. 
However, the Shew \& Lathrop experiment remains to date the only rotating convection experiment running with liquid sodium. 
Measurements rely on thermocouples immersed within the fluid, and velocities are obtained by correlating neighbouring thermocouples and using Taylor's hypothesis.
They focussed on the turbulent regime and observed 
small-scale convective motions with a strong, large-scale retrograde azimuthal flow. 
The local intensity of the zonal flow intensity fits a scaling of $u_\theta\simeq 3.5 \Omega\alpha D\Delta T$ obtained from a simple balance between Buoyancy and the Coriolis force. 
As one would expect, they observe developed turbulence with weak convective heat transport and a near-diffusive temperature profile. 
They measured Reynolds numbers in the range $Re\sim 10^3-10^4$,
and a total radial heat flux following a $Nu\sim Ek^{-1/3} Ra$ scaling, indicating a rotation-dominated heat transfer \cite{kunnen2021_jot, ecke2023_arfm}. Extrapolating these measurements to Earth leads to an estimated Rayleigh number of $Ra\sim 10^{23}$ and convective velocities comparable to the azimuthal flow of $\simeq 2\times10^{-4}$ m/s, from which they estimated the magnetic Reynolds number in the equatorial regions of the Earth to be $Rm\simeq2\times10^2$.
They identified a "knee" in the energy spectra where energy was injected by convection into the flow, which, extrapolated to Earth, corresponds to a turnover time of 30 days, and a size of \SI{1}{km}, again well below any observable scale. They also derived an estimate for the Joule dissipation {in the Earth's core} in the range of \SI{10}{GW} to \SI{10}{TW}.
\subsection{Rotating magnetoconvection in spherical annulus experiments}
\label{sec:sphere_mhd}
Extrapolating rotating convection experiments to Earth yields estimates that are inconsistent with geomagnetic data, for which the Lorentz force usually takes the blame. 
It was therefore natural for rotating magnetoconvection experiments to emerge. 
There are essentially three strategies to go about these experiments: letting the fluid motion itself generate its own magnetic field, putting a magnet around the experiment, or bringing the experiment to a magnet.

The first strategy, dynamo experiments, however, poses such a challenge on its own that no working concept of a convective dynamo experiment has been put forward yet. There are even claims that such experiments are impossible at laboratory scale \cite{Tilgner2000}. {Time will tell...}

\begin{figure}[h!]
\includegraphics[width=\textwidth]{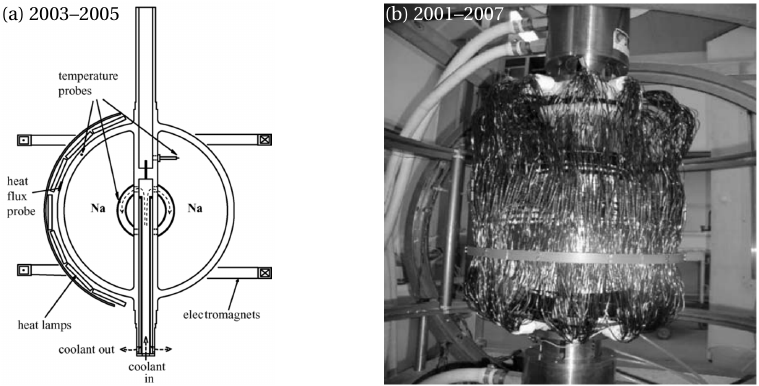}
\caption{Liquid metal experiments:
(a) 2003--2005: Shew \& Lathrop's sodium experiment with a Helmholtz coil generating an axial magnetic field \cite{shew2005_pepi}
(b) 2001--2007: Gillet et al's Gallium experiment with their "hairy magnet" made of a long toroidal wire wrapped around the shell, generating an azimuthal magnetic field of up to \SI{0.03}{T}\cite{gillet2007_jfm2}.\label{fig:spheremhd_setup}}
\end{figure}

The second strategy has been pursued to date by only two magnetoconvection experiments with a spherical geometry by Shew \& Lathrop and Gillet et al. \cite{shew2005_pepi,gillet2007_jfm2}, both represented on figure \ref{fig:spheremhd_setup}. 
These experiments rely on the rotating convection devices previously built by the Maryland and Grenoble groups \cite{shew2005_pepi,aubert2001_pepi,gillet2007_jfm2}. The main drawback of this strategy is that only relatively weak magnets can be transported and fitted around existing setups, and thus, deliver limited magnetic fields.
In the geophysical context, the Lorentz force is typically assessed relative to the Coriolis force by the Elsasser number $\Lambda=\sigma B_0^2/(2\rho \Omega)$, which for the Earth is deemed to lie between 0.1 and 100 \cite{cardin1994_pepi,schubert2011_pepi,roberts2013_rpp,horn2022_prsa,aurnou2015_pepi}. 
Shew \& Lathrop \cite{shew2005_pepi} placed a \SI{3}{mT} Helmholtz coil around their spherical annulus providing $\Lambda\leq1.9\times10^{-4}$ and saw a global decrease in heat flux with increasing magnetic field.
Gillet and collaborators \cite{gillet2007_jfm2}, created a "hairy magnet", see figure \ref{fig:spheremhd_setup}, by wrapping 444 turns of a wire around meridional planes and through the central shaft of their spherical annulus 
to generate an azimuthal field of up to \SI{0.03}{T}, and $\Lambda\leq9.94\times10^{-2}$.
They found a zonal flow following the non-magnetic phenomenology \cite{gillet2007_jfm2},
but with larger azimuthal length scales and magnetically controlled zonal velocity scaling as $\bar U\sim (\widetilde U_r \widetilde U_\theta)^{2/3}\sim \widetilde U_r (l_\theta/l_\beta)^{2/3}$.
Hence, both experiments showed the stabilising effect predicted at low $\Lambda$ \cite{fearn1979_prsa,soward1979_pepi}. 
Their relevance to the Earth is, however, questionable considering the Lorentz force may become destabilising at greater values of $\Lambda$ around unity \cite{Chandrasekhar1961,eltayeb1972_prsa,eltayeb1975_jfm,fearn1979_prsa,aujogue2015_pf,horn2022_prsa}. 

Finally, the third strategy, and the power of large magnets it affords, has only been exploited in set-ups targeting the polar region, namely LEE1 and LEE2, which took advantage of fields up to \SI{10}{T}. These are discussed in section~\ref{sec:TC}.
\subsection{Experiments with inhomogeneous heating}
In most experiments, great efforts are 
deployed to keep homogeneous temperature boundary conditions, so as to preserve the
ideal character and to make the results universal. Yet, there is evidence that 
large-scale magnetic flux anomalies are driven by large-scale inhomogeneities at the 
CMB \cite{gubbins1986_grl,bloxham1987_nat,johnson1998_grl} and 
that they may drive magnetic field reversals {\cite{glatzmaier1999_nat,kutzner2004_grl,wicht2010_ssr,olson2014_pepi,wicht2016_pepi,terranova2024_sr,tarduno2015_natcomm}}. Heat
flux heterogeneity at the CMB may thus play a role in the interior dynamics. 
Hart \cite{hart1996_3mg, Hart1999} made the first attempt at studying them but suffered from limitations inherent to space experiments, when the heat source failed, as discussed in section \ref{sec:central}. Since then, two experiments studied their effect on rotating convection systematically.

\begin{figure}[h!]
\includegraphics[width=\textwidth]{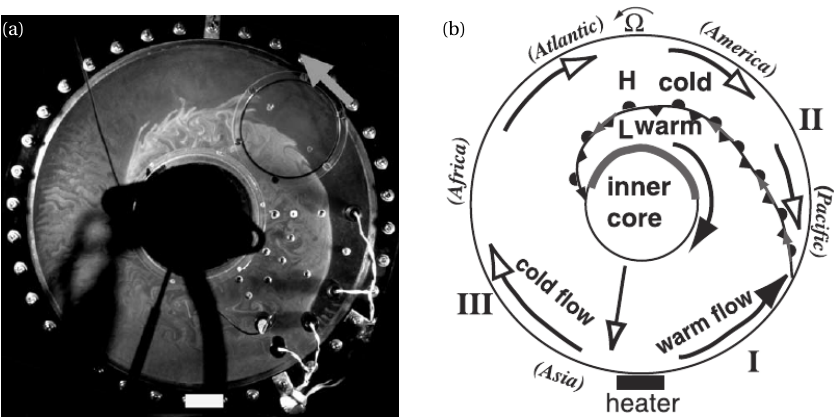}
\caption{Flow patterns produced by localised heating of the outer boundary of a hemispherical shell \cite{sumita2002_jgr}. (a) Inner core spiral developing in the global locking mechanism as a result of localised heating at the outer hemisphere, visualised by injection of dye \cite{sumita2002_jgr}. (b) the front forming the spiral results from the collision of the eastward and westward jets originating at the heating anomaly, a mechanism suspected to take place in the Earth's core (right). \label{fig:spiral}}
\end{figure}

Sumita \& Olson \cite{sumita1999_sci,sumita2002_jgr} adapted their experiment \cite{sumita2000_pepi,sumita2003_jfm}
and created a local thermal inhomogeneity with a small rectangular heater attached to the outer boundary. Its size and latitudinal position were varied. This extra heating models a local cold anomaly of the Earth's CMB where the outward heat flux locally is higher since the temperature gradient in the experiment is reversed compared to the Earth's outer core.
The total heat flux was monitored by comparing the inlet and outlet temperatures of the inner sphere's cooling shaft. The intensity of the anomaly was quantified by the ratio of excess heat flux at the rectangular heater $q_h$ to the flux at the inner boundary $Q$, $Q^*=q_h/Q$.
The heterogeneity drives a local prograde (eastwards) flow. Two different regimes exist: For $Q^*< 0.7$, in the so-called \emph{local locking} regime, this flow remains localised near the outer boundary. For stronger inhomogeneities, in the \emph{global locking} regime, a large-scale spiral with a sharp front develops across the whole gap, see figure \ref{fig:spiral}. The front separates warm and cold regions and is accompanied by a thin jet linking the outer and inner boundaries, which is responsible for an increase in global heat flux and large inhomogeneities at the inner boundary.
The regimes are practically independent of the size of the heater but depend on its latitudinal position: at low latitudes, the prograde flow forms but does not develop into a front, whereas at higher latitudes, the front enters a regime of periodic formation and destruction.
Sumita \& Olson estimated the heat flux beneath East Asia to be $Q^*\simeq2$, based on the CMB temperature and its variations \cite{boehler2000_rg,castle2000_jgr,garnero2000_areps} and seismic models \cite{vanderhilst1999_sci}. For this value, global locking is possible, so they proposed that \emph{"the existence of this front in the core may explain the Pacific quiet zone in the secular variation of the geomagnetic field and the longitudinally heterogeneous structure of the solid inner core."}

Sahoo and Sreenivasan \cite{sahoo2020_epsl,sahoo2020_jfm} built a cylindrical annulus filled with water (height \SI{37}{cm}, inner/outer radii \SI{5}{cm}/\SI{14.2}{cm}), with flat, thermally insulating top and bottom boundaries, mostly radial centrifugal force and PIV in horizontal planes \cite{sahoo2020_epsl, sahoo2020_jfm}.
At the inner boundary the temperature is imposed, while at the outer boundary, the heat flux is imposed by the heater fitted around the annulus.
Unlike Sumita \& Olson's experiment, the inhomogeneity is not local but takes the form of an azimuthal variation of boundary heat flux controlled by dividing the external heater into four independently controlled $\pi/2$ sectors.
The system spanned inhomogeneities in the range $0\leq Q^*\leq2$. Here $q_h$ is the amplitude of the azimuthal flux variation.

\begin{figure}[h!]
\includegraphics[width=0.91\textwidth]{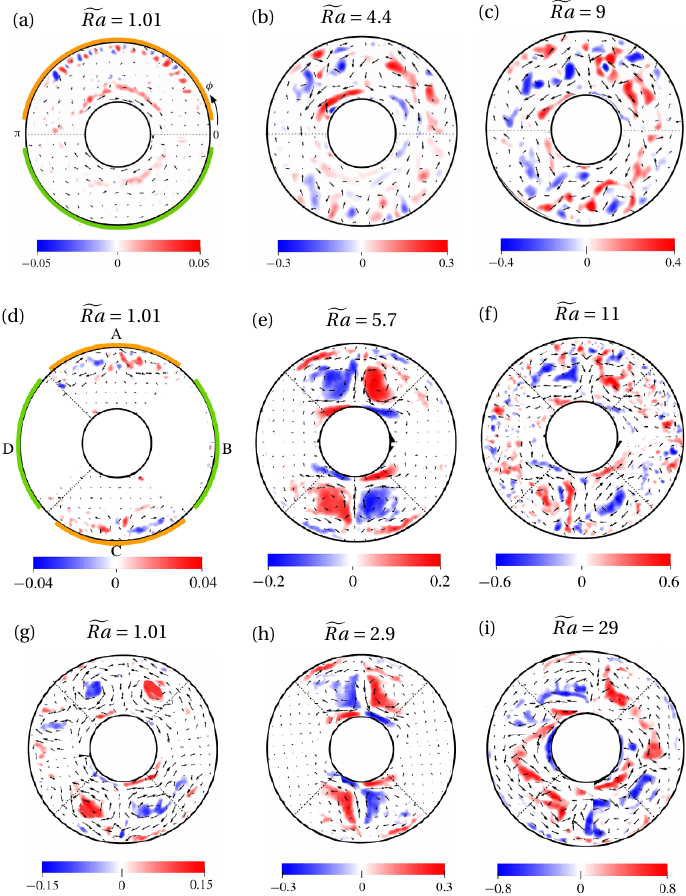}
\caption{Horizontal velocity vectors (arrows) and shaded contours of axial vorticity (s$^{-1}$) in horizontal planes of {Sahoo \& Sreenivasan's experiment} \cite{sahoo2020_jfm} in an annulus with a heated outer boundary with two or four sectors of independently adjustable heat flux (yellow and green colours indicate regions or high and low heat flux respectively). Plots are averaged in time, and shown for different levels of supercriticality $\widetilde{Ra}$. (a)-(c): one-fold variations with $Q^*=0.7$, (d)-(f): two-fold variation at $Q^*=1$ showing regionalisation of convection, (g)-(i): two-fold variation with $Q^*=2$, homogenisation of convection takes place at sufficiently high forcing under the effect of the azimuthal flow. \label{fig:sahoo2020_piv}}
\end{figure}

Experiments were conducted with alternating high and low fluxes every $\pi$ (one-fold symmetry) and every $\pi/2$ (two-fold symmetry). Both configurations lead to a slight reduction of the critical flux-based Rayleigh number for the onset of convection. Examples of vorticity fields are represented on figure \ref{fig:sahoo2020_piv}.
 The flow consists of counter-rotating vortex pairs separated by thin downwellings similar to Sumita \& Olson. These large scales are accompanied by small-scale motion. With a two-fold variation, the flow becomes homogeneous above about 30 times the critical Rayleigh number, most likely as a result of the strong azimuthal flow that exists in this regime. Homogenisation was not observed with a one-fold variation as convection in the low-flux sector is not able to transport motion across it. Sahoo \& Sreenivasan interpreted the magnetic field inhomogeneities as signatures of CMB inhomogeneities to estimate that $Q*>2$ for the Earth, implying regions of subadiabatic heat flux.
Achieving these in the experiments, would, however, require adding active cooling to the outer wall. Nevertheless, such inhomogeneities in the Earth's magnetic field suggest that homogenisation does not take place in the equatorial regions of the core.
This offers a potential method to constrain the Rayleigh number in the Earth's outer core. 
\section{The default setup for axial gravity: cylinders \label{sec:axial}}
Experiments in geometries where the natural terrestrial gravity, the temperature gradient and the imposed rotation are aligned (fig.~\ref{fig:geometries} (c)) capture the part of the planet that the experiments with cylindrical radial gravity neglected - the polar region.
The default set-up is a straight cylinder, the shape that naturally arises in rapidly rotating systems. Only three experiments were conducted in box geometries \cite{Sakai1997, Boubnov1986, Fernando1991} which creates the undesired issue of developing corner flows that have no planetary equivalent. These Cartesian geometries, however, ease visual access in experiments.

Convection in rotating spherical convection onsets in the equatorial region and thus was initially thought to be the most relevant for planetary interiors. Nowadays, it is, however, clear that the polar region is at least equally important in the more supercritical regimes where planets reside \cite{Gastine2023, Schaeffer2017}.
Thus, the importance of some of the classical rotating convection experiments with axial gravity to geophysical processes is sometimes only emphasised more strongly in hindsight \cite{horn2022_prsa, Horn2024, roberts2013_rpp, King2013, King2015}. Nonetheless, as visualised in fig.~\ref{fig:timeline} 
, this configuration has been the most common and enduring one since the beginning of rotating convection experiments and shows no signs of losing its popularity. There are several recent thorough reviews detailing the experiments, their results and more \cite{cheng2018_gafd, aurnou2015_pepi, kunnen2021_jot, ecke2023_arfm} so that here we keep the discussion very brief.

\begin{figure}
\includegraphics[width=\textwidth]{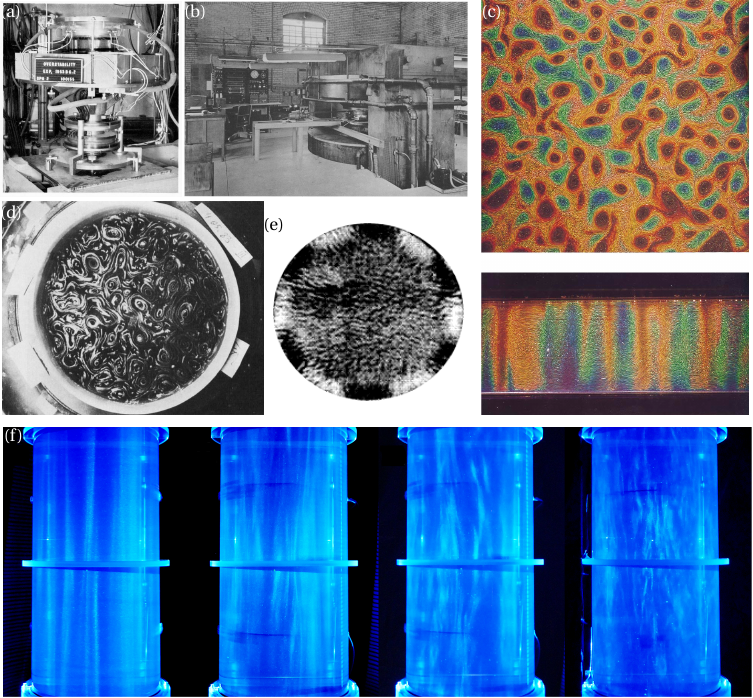}
 \caption{(a) The first experiment of rotating convection in Chicago on the 20 November 1953 \cite{Fultz1954}. (b) The hydromagnetic laboratory at the University of Chicago where Nakagawa did his seminal rotating convection experiments. \cite{Chandrasekhar1957} (c) Visualisation of columns in rotating convection in water using thermotropic liquid-crystal capsules by Sakai, top and side view (top/bottom) \cite{Sakai1997}. (d) Visualisation of chaotic vortices in rotating convection in water using aluminium powder on the top surface by Boubnov and Golitsyn \cite{Boubnov1986}. (e) Visualisation of wall modes in water using shadowgraph by Zhong et al. \cite{Zhong1991}. (f) Visualisation of the flow regimes in turbulent rotating convection in water using rheoscopic particles and illumination with a vertical light sheet; left to right: convective Taylor columns, plumes, geostrophic turbulence, rotationally influenced turbulence by Cheng et al. \cite{Cheng2020} \label{fig:cylexp}}
\end{figure}

Arguably the first systematic and quantitative experiments of rotating convection was a series of mercury experiments conducted by Fultz, Nakagawa, Frenzen and Goroff in the years 1953 to 1960 at the University of Chicago \cite{Fultz1954, Fultz1955, Nakagawa1955, Nakagawa1957, Nakagawa1959, Goroff1960}, the first one conducted on the 20 November 1953 as shown in fig.~\ref{fig:cylexp}(a). (A few earlier rotating experiments were also conducted by Nakagawa and Fultz at the University of Tokyo and Chicago, respectively, in water and air, however, the upper surface was cooled by evaporation which is a slightly different fluid dynamics problem from the Rayleigh-B\'enard-like set-up we consider throughout this review \cite{Nakagawa1955}.) These seminal experiments were very comprehensive. Since they were conducted in a liquid metal with $Pr \approx 0.025$, oscillatory convection was naturally included. Moreover, Nakagawa had access to an electromagnet of a discarded \SI{36.5}{in} cyclotron which had been reconditioned at the Enrico Fermi Institute for Nuclear Studies at the University of Chicago with a field strength of up to \SI{1.3}{T}, shown in fig.~\ref{fig:cylexp}(b) \cite{Chandrasekhar1957}. This allowed him to also study rotating magnetoconvection. These experiments were mainly concerned with the flow behaviour close to the onset of convection and the validation of linear stability results derived by Chandrasekhar and Elbert \cite{Chandrasekhar1961, horn2022_prsa, Horn2024}.

The focus on relatively low supercriticalities remained until the end of the last century. The most common working fluid was water \cite{Rossby1969, Boubnov1986, Boubnov1990, Fernando1991, Liu1997, Zhong1991, Zhong1993, Ning1993,  Vorobieff1998, Sakai1997} which allows direct optical access and visualisation, but prohibits oscillatory convection. 
The same applies to silicone oil as tested by Koschmieder \cite{Koschmieder1967} and Rossby \cite{Rossby1969}. These earlier experiments established the formation of regular cell patterns and convective Taylor columns as the typical structures of steady bulk convection, see fig.~\ref{fig:cylexp}(c), but also found irregular vortex patterns; an example is shown fig.~\ref{fig:cylexp}(d).
Also, the heat transport and its scaling with the control parameters were investigated, especially in the few visually opaque experiments (either due to the fluid or by design) by Rossby in mercury \cite{Rossby1969}, Donnelly and collaborators in cryogenic helium-I \cite{Lucas1983, Pfotenhauer1984, Pfotenhauer1987}, and Aurnou and Olson in gallium \cite{Aurnou2001}. None of the experiments was likely extreme enough to capture the geostrophic turbulent heat transport scaling one may expect in Earth's outer core or planetary interior, though.

Likely the most crucial finding in that period, however, was the discovery of wall modes by Ecke and collaborators, shown in fig.~\ref{fig:cylexp}(e) \cite{Zhong1991, Zhong1993, Ecke2023b}. Wall modes explained why the onset of convection as determined e.g. by heat transport measurements did not agree with Chandrasekhar's predictions but was indeed at lower $Ra$. The destabilising effect of the wall and the breaking of the symmetry results in structures that travel along the periphery of the rotating convection vessel and exponentially decay towards its centre. Initially thought to be mainly an onset phenomenon and an experimental restriction, wall modes are now known to be much more persistent and transform non-linearly into a boundary-zonal flow affecting also strongly supercritical settings \cite{Zhang2020, Wedi2022, Zhang2021, Shishkina2020, Vogt2021, Favier2020, deWit2023, Terrien2023}. Moreover, they may also exist along the virtual boundary of the tangent cylinder in Earth's outer core, as we will discuss in the next section~\ref{sec:TC}.

In our current 21\textsuperscript{st} century, the goal has changed and is now rather to push turbulent rotating convection to the extreme, such as in the TROCONVEX experiment in fig.\ref{fig:cylexp}(f). \cite{kunnen2021_jot, Cheng2020}. The most popular working fluid is still water \cite{King2009, Cheng2015, Liu2009,Cheng2020, Zhong2009, Weiss2010, Weiss2011, Kunnen2008, Kunnen2010, Rajaei2016,Joshi2016, Rajaei2017, cheng2018_gafd, Cheng2020,  Kunnen2021, Ding2021, Lu2021, Hu2022, Shi2020}, but also cryogenic helium \cite{Niemela2010, Ecke2014}, silicone oils  \cite{Abbate2023}, sucrose solution \cite{King2009}, gallium \cite{ Aurnou2018, King2013, Vogt2021, Grannan2022}, and pressurised SF\textsubscript{6} \cite{Zhang2020,Wedi2021}.
The connection to geophysics and planetary physics is sought much more explicitly, in particular, in the Calimero (Californian Model of Earth's Rotation), NoMag, RoMag (all three in Los Angeles, Aurnou and collaborators \cite{Abbate2023, Hawkins2023, King2009}) and TROCONVEX (Eindhoven, Kunnen and collaborators \cite{cheng2018_gafd}) experiments. Notably, RoMag, is the only operative liquid metal rotating magnetoconvection experiment with thus, the closest resemblance to planetary core fluid.
Thus, this meant establishing the possible regimes of geostrophic convection (see fig.\ref{fig:cylexp}(e), convective Taylor columns, plumes, geostrophic turbulence, rotationally influenced turbulence), testing various scaling for the heat and momentum transport, length scales, the importance of centrifugal effects and much more.
Several of these experiments in China, the USA and The Netherlands, are still active and running and producing results (see fig.~\ref{fig:timeline}), thus, we may expect more to come that will enrich our understanding of the polar physics in planetary interiors in the future.

\section{Tangent Cylinder dynamics \label{sec:TC}}
Rotating convection in cylinders, reviewed in section~\ref{sec:axial}, may represent the dynamics in the polar regions. However, those experiments do likely not fully capture the entirety of the polar region that extends up to the virtual tangent cylinder (TC) boundary, see fig.~\ref{fig:geometries} (a). The sidewall boundary conditions in cylindrical experiments are usually impermeable and adiabatic. The boundary conditions of planetary tangent cylinders are, however, likely neither. The most recent data from the SWARM mission clearly shows a strong jet wrapped around the TC and, importantly, meandering in and out of the TC \cite{finlay2023_nat}. Similarly, dynamo simulations at extreme enough parameters show flows across the TC \cite{Schaeffer2017}.

Thus, several major questions arise: First, under which conditions can the \emph{Taylor-Proudman constraint} (TPC, see section~\ref{sec:phys}) be violated such that the TC boundary becomes permeable? Second, how much heat is transferred between the polar and equatorial regions across the TC? Finally, given such drastic differences in boundary conditions, how well do the experiments in cylinders truly represent polar convection?
\subsection{Aurnou et al.'s TC: What's new in Baltimore?}
\begin{figure}
\includegraphics[width=\textwidth]{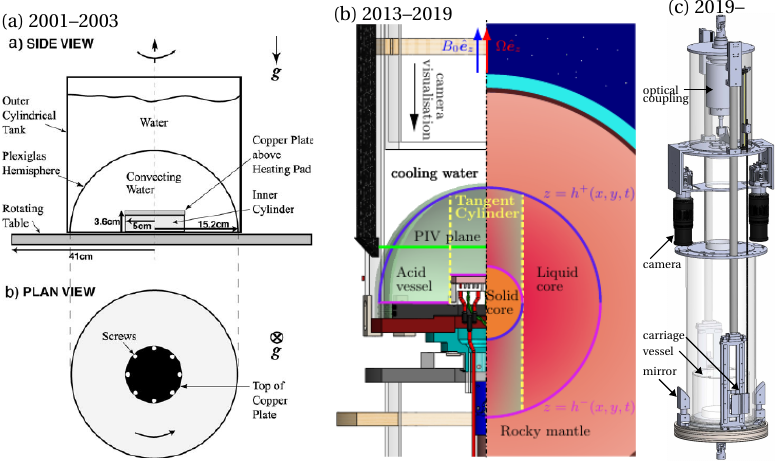}
\caption{Sketches of the three experiments reproducing the TC geometry:
(a) 2001--2003: Rotating vessel of Aurnou et al. showing the puck-shaped heater inside the hemispherical dome filled with water \cite{aurnou2003_epsl} (b) 2013-2019: Rotating part and static support of LEE1 set against a section of the Earth illustrating how the inner core, ICB and CMB are modelled in the experiment \cite{aujogue2016_rsi,aujogue2018_jfm,potherat2023_prl} (c) 2019--: Rotating part of LEE2, with cameras, mirrors and optical slip ring, where the fluid vessel is now cylindrical instead of hemispherical. The two large black cameras near the top and the two mirrors fitted at the bottom are part of the PIV system for vertical planes. The vertically travelling carriage near the bottom holds the lens generating the horizontal LASER plane for PIV in horizontal planes of adjustable height \cite{agrawal2024_gji}.
\label{fig:tc_setups}}
\end{figure}

Aurnou and collaborators \cite{aurnou2003_epsl} undertook the first attempt at addressing these questions in their Baltimore lab. They built a setup with a hemispherical vessel (\SI{15.2}{cm} diameter) filled with water, sitting in a cylindrical bath also filled with water ensuring a cold, approximately homogeneous temperature at its outer boundary (figure \ref{fig:tc_setups}(a)). A heater shaped like a hockey puck (\SI{10}{cm}-diameter, \SI{3.6}{cm} high) rested at the flat bottom of the hemisphere. 
The heater achieved two functions: First, it delivered a constant heat flux $q$ into the fluid. Second, the outer edge of the puck formed a discontinuity in the height of the fluid domain in the same way as a planetary {solid core} does. The choice of a puck shape for the heater, instead of a spherical one was motivated by the difference in gravitational fields between the Earth and the experiment: in the experiment, the axial gravity would misalign with a spherical boundary of near-constant temperature and drive unwanted baroclinic flows in its vicinity. By contrast, the flat upper surface of the heater is everywhere perpendicular to the gravity, and so eliminates that problem. Hence, in the lab, pucks are a better representation of the Earth's inner core than spheres.

\begin{figure}
%
\includegraphics[width=\textwidth]{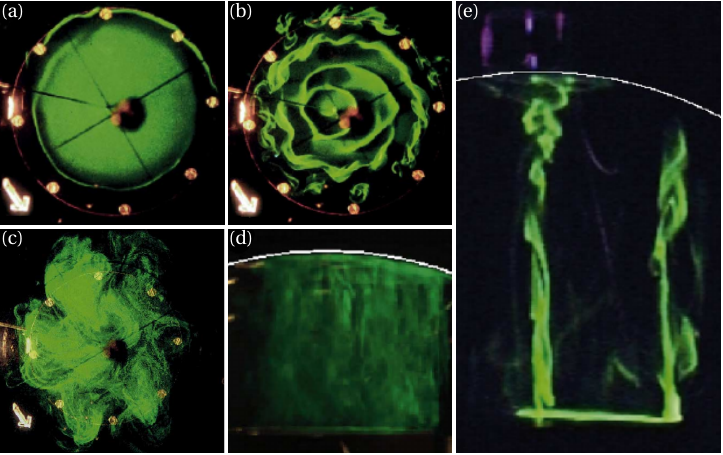}
\caption{Dye visualisations showing the regimes of rotating convection inside a TC during the instationary phase following the ignition of thermal forcing from a state of solid body rotation by Aurnou et al. \cite{aurnou2003_epsl}.
	Top view of (a) the onset of convection and  (b) helical plumes at  $Ra = 9.01\times10^9$ ; $Ek = 9.65\times10^{-5}$,
	(c) side view of  baroclinic instability developing across the TC boundary,
	(d) side view of fully developed convection inside the TC before the onset of baroclinic instability at ($Ra = 1.10\times10^{10}$,  $Ek = 4.11\times10^{-5}$). Arrows indicate the direction of rotation. (e) Plumes arising near the TC boundary at the onset of convection ($Ra = 4.44\times10^9$, $Ek = 4.26\times10^{-5}$).
\label{fig:tc_hemisphere}}
\end{figure}

Aurnou and collaborators used dye visualisation and thermistors embedded near the heater surface to measure heat fluxes and velocities.
They identified five successive regimes based on the flow structure observed as the thermal forcing ramps up during a transient experiment. These regimes map to those in cylindrical vessels but for specificities inherent to the TC geometry. At low thermal forcing,  
no convection is detected inside or outside the TC but non-vertical residual thermal gradients drive a very weak thermal wind that suppresses convection and thus increases the critical Rayleigh number for the onset of convection inside the TC. Increasing the thermal forcing leads to a ring of vortices attached to the outside of the TC. These vortices result from baroclinic instability and remain confined outside the TC due to the TPC.
Inside the TC, convection is still absent. At higher thermal forcing, convection sets in within the TC at a slightly higher critical Rayleigh number than in cylindrical vessels with adiabatic sidewalls because of the weak thermal wind. The onset structures are helical, prograde near the heater and retrograde near the top. They ignite near the TC boundary and their motion is confined within the TC by the TPC as visualised during a transient experiment in figure \ref{fig:tc_hemisphere}(b). Further increasing the thermal forcing, but keeping strong rotation, the helical structures fill the entire TC, in a regime that would nowadays be referred to as "rotation-dominated columnar regime" \cite{kunnen2021_jot,Julien2012}. When rotation is less dominant over buoyancy, 
the flow structures do not extend over the full height of the volume 
but remain strongly affected by rotation \cite{kunnen2021_jot,Julien2012}. 

The presence of baroclinicity in background rotation drives a strong azimuthal wind with maximum velocity $U_\phi^{\rm max}$ near the inner TC boundary. The scaling of its intensity with the convective buoyancy flux $q_B=\alpha g q/(\rho C_p)$, as $U_\phi^{\rm max}\sim 2.05(q_B/\Omega)^{0.52\pm 0.03}$ confirms this mechanism. The authors suggest that this mechanism and its confinement to polar regions within the TC may explain the polar vortices inferred there from the 1980 Magsat and 2000 Oersted satellite missions\cite{hulot2002_nat}. Extrapolation of their experimental observation to Earth leads to estimates of a few km for the diameter of columnar vortices within the TC and a flux Rayleigh number of the order of $10^{31}$. They do stress however that testing this thermal wind scaling for the Earth would require incorporating such effects as the Lorentz force, compressibility and other possible effects.

\subsection{Little Earth Experiment I (LEE1)}
Poth\'erat and collaborators used design ideas similar to Aurnou et al. \cite{aurnou2003_epsl} in the \emph{Little Earth Experiment} (LEE1), a hemispherical dome of diameter \SI{28.5}{cm}, and a puck-shaped heater \SI{2.25}{cm} in height and \SI{10}{cm} in diameter. They introduced two novelties with drastic consequences for the experimental realisation (figure \ref{fig:tc_setups}(b)). First, they incorporated PIV measurements to capture the detail of the time-dependent velocity field in horizontal planes and in one meridional plane. Second, LEE1 targeted magnetostrophic regimes with an Elsasser number of $\Lambda\sim1$ \cite{aujogue2016_rsi, aujogue2018_jfm, potherat2023_prl}. 

Combining optical methods and strong MHD effects requires a working fluid that is both transparent and electrically conducting. The fluid with the highest electrical conductivity satisfying these constraints is sulphuric acid at 30\% mass concentration, of conductivity $\sigma=$\SI{86}{S/m}, i.e. $10^4$ times less than a typical liquid metal. Luckily, sulphuric acid is also the least dangerous amongst potential candidates (hydrofluoric acid, ionic liquids etc.). 
Since sulphuric acid is typically 5 to 10 times less dense than a liquid metal, it is easy to see that to reach $\Lambda\sim1$ at $Ek\sim10^{-6}$, magnetic fields of the order of \SI{10}{T} are needed. To complicate matters, the magnetic field must pervade a sufficiently large volume to host a fluid dynamics experiment with its instrumentation i.e. a cylindrical volume of about \SI{30}{cm} diameter over about \SI{1}{m}. At the time of LEE1's operation, only one magnet in the world was capable of delivering such a magnetic field: the \SI{12}{MW} M10 magnet of the Grenoble High Magnetic Field Laboratory (LNCMI-G) with up to \SI{10}{T} in a \SI{376}{mm}-diameter/\SI{2}{m}-long warm bore (Nowadays, LNCMI-G offers up to \SI{18}{T} in their new hybrid magnet \cite{pugnat2022_ieee}). Using such a large magnet meant that only one strategy was reasonably possible for the design of LEE1 out of the three outlined in section \ref{sec:sphere_mhd}: LEE1 had to be built to fit into the existing magnet. This imposed further severe constraints on the design. Materials rotating within the bore had to be non-conducting to avoid Foucault currents, and those in contact with the working fluid had to possess long-term resistance to highly concentrated acids. Electronic equipment and electric motors cannot operate in high magnetic fields and have to be deported either far above or far below the magnet. The space between the vessel and the inner wall of the magnet bore is only a few cm. Thus, visualisation from the side had to rely on small concave mirrors providing sufficient angular access to capture the entire height of the TC. Finally, LEE1's motor was located about \SI{0.5}{m} underneath the bottom of the magnet. The acquisition system (camera, laptop controlling PIV LASERS and recording thermocouple signals) was placed on a rotating platform about \SI{2}{m} above the top of the magnet. 
In total, LEE1 was approximately \SI{4.5}{m} in height and linked by a structure mostly made of different plastics.

The heater was made of a plastic heat exchanger circulated by ethylene-glycol heated in the static frame. The top of the heater delivering heat to the acid was made of SHAPAL, a heat-conducting, electrically insulating and acid-resistant ceramic. 
This design avoided electric parts but its thermal inertia prevented any feedback control of the temperature difference.

\begin{figure}
\includegraphics[width=\textwidth]{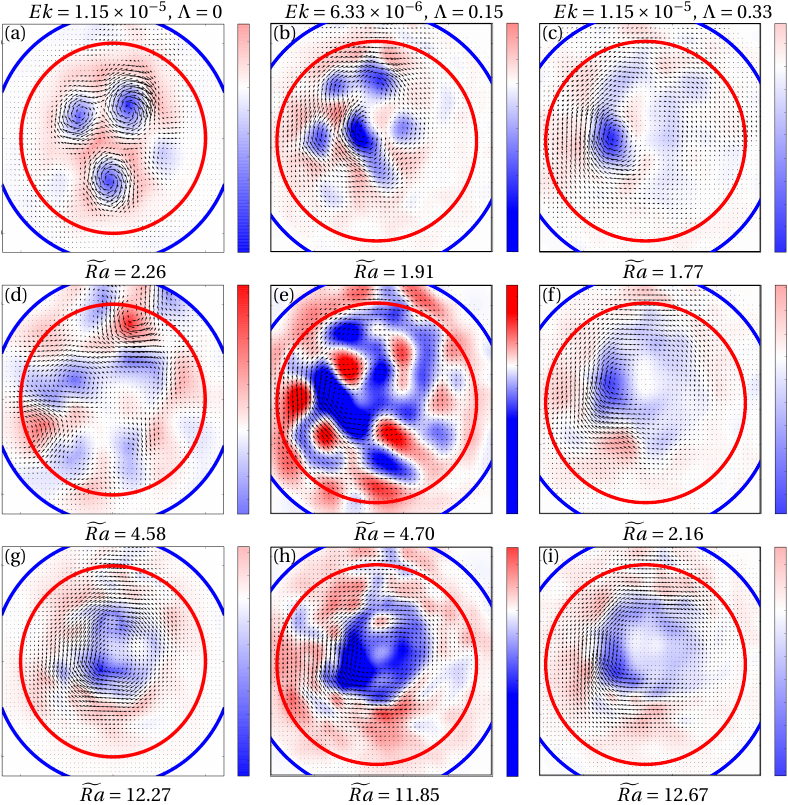}
\caption{PIV visualisation of the non-magnetic and MHD flow patterns inside the TC in LEE1 \cite{aujogue2016_phd,aujogue2018_jfm} showing vorticity (Blue denotes negative vorticity, red positive vorticity.) and velocity (arrows) averaged in time in a plane located at 51$^\circ$ latitude (\emph{i.e.} just below the point where the TC meets the hemispherical vessel wall). Values of $\widetilde{Ra}$ are calculated with respect to the critical Rayleigh number for $\Lambda=0$, even for cases where $\Lambda>0$. The red circle indicates the position of the TC boundary whereas the blue circle corresponds to the vessel wall.
The non-magnetic case (a,d,g) exhibits large, merged columns in the weakly supercritical regime. For higher criticality, wall modes are visible and for the highest criticality, a large central retrograde vortex exists at the centre.
At $\Lambda=0.15$ (b,e,h), the effect of the magnetic field is not pronounced and the flow patterns resemble those at $\Lambda=0$.
For $\Lambda=0.33$ (c,f,i), a large {off-centre} retrograde structure forms a much lower criticality, with a zero-vorticity "eye" in its centre.
\label{fig:lee1_piv}}
\end{figure}

\paragraph{Non-magnetic rotating convection}
LEE1 was first operated in water to focus on non-magnetic rotating convection in the TC. In the entire range of parameters investigated, the flow was mostly confined within the TC, i.e. the TPC was sufficiently well satisfied for the TC boundary to be practically impermeable, 
with only some weak motion detected outside the TC, see figure \ref{fig:lee1_piv} (a,d,g).
The onset of convection follows similar laws to the onset of planar layer convection, but with slightly different prefactors: close for the critical Rayleigh number $Ra_c = (32.3 \pm 4) Ek^{-1.29\pm0.05}$, but lower for the critical wavenumber $k_c=(0.58\pm 0.08) Ek^{-0.32\pm0.05}$. Past the onset, both wall modes and bulk modes 
are observed, which implies that the TC boundary acts as a solid impermeable wall in conditions relatively close to onset \cite{goldstein1993_jfm}. Unlike in cylindrical tanks, however, wall modes were not detected below the onset of bulk convection. A possible explanation is that the TPC does not erect a boundary when the flow is still. In other words, below the onset of convection, the TC does not exist as a wall, so neither do wall modes.
With increasing supercriticality, the flow follows a complex sequence of patterns to arrive at a state dominated by a single retrograde vortex for $\widetilde Ra\simeq 10$.
At high latitude, this vortex is associated with a retrograde thermal wind located \emph{inside} the TC, whose intensity follows a scaling nearly identical to that found by Aurnou et al. \cite{aurnou2003_epsl} $Ro = (5.33\pm 0.3) (Nu\, Ra\, Ek^3 Pr^{-2})^{0.51\pm0.04}$.
The heat flux shows a clear transition between regimes of rotation-dominated convection scaling as $Nu\simeq 0.38 Ek^2 Ra^{1.58}$ and buoyancy-dominated convection with $Nu\simeq 0.2 Ek^{0.33}$ which resembles convection in cylinders and plane layers \cite{kunnen2021_jot,ecke2023_arfm}.

\paragraph{Magneto-rotating convection\label{sec:tcmhd}}
Unlike liquid planetary cores, LEE1 operates in the quasi-static MHD regime 
where the feedback of the flow onto the magnetic field is negligible, so the magnetic field is imposed, and constant, and dynamo processes are excluded \cite{roberts1967}. The acid's low conductivity also precludes the occurrence of Alfv\'en waves or magneto-Coriolis waves, even though these can exist a low magnetic Reynolds numbers in liquid metals \cite{schmitt2008_jfm,lalloz2024_jfm} and bear relevance to those in planetary cores \cite{gillet2010_nat,gillet2022_pnas}. Hence, LEE1 focuses on capturing the feedback of the magnetic field on convection rather than the reverse. 

The magnetic regimes differ from the nonmagnetic ones in a number of ways, see figure \ref{fig:lee1_piv}. At the most fundamental level, the Coriolis force now competes with the Lorentz force, so the TPC no longer applies. Instead, a \emph{Magnetic Taylor-Proudman constraint} (MTPC) imposes a kinematic relation between the flow along the TC boundary ($u_\phi$) and the flow through it ($u_r$):
\begin{equation}
\Delta_{rz}\langle u_r\rangle_{t,\phi}+ \Lambda\partial^2_{zz}\langle u_\phi\rangle_{t,\phi}=0.
\label{eq:kinematic_mpt_rtheta}
\end{equation}
This theory is confirmed using PIV in two horizontal planes at low and high latitudes showing that a meridional flow through the TC boundary follows the MTPC \cite{potherat2023_prl}. As a result of this reorganisation of the flow, the azimuthal flow does not follow the thermal wind scaling and increases significantly with $\Lambda$.
Applying the MTPC to the Earth would require a more general form to incorporate the complex geometry of the Earth's magnetic field. Nonetheless, the MTPC provides at least a plausible mechanism to account for the meandering of the zonal flow in and out of the TC inferred from the SWARM data \cite{finlay2023_nat}, and previous simulations suggesting that the TPC was broken at the TC boundary \cite{Schaeffer2017,cao2018_pnas}.

The main effect of the Lorentz force inside the TC is a suppression of heat transfer. At $\Lambda>0.2$, Nusselt numbers are systematically much lower than their counterparts at $\Lambda=0$, but scale with a greater power of $Ra$ (up to 1). Unlike in the non-magnetic case, the highest thermal forcing available, $Ra\sim 10^{9}$, is insufficient to reach a transition to a buoyancy-dominated regime \cite{aujogue2016_phd}. {This does not imply, however, that the presence of a magnetic field always suppresses heat transfer in general. In spherical shells for example, dynamo simulations at $Pr=1$ show enhancement of heat transfer at $Ek\lesssim10^{-5}$ \cite{yadav2016_gafd} and the trend observed in LEE1 does not exclude that the imposed magnetic field could lead to heat transfer enhancement in the TC at higher Rayleigh numbers than accessible in the experiment. This just goes to show how much is left to understand the basics of magneto-rotating convection.}
Detecting the onset of rotating magneto-convection inside the TC would require far more magnet time than possible in M10 but several states of convection are detected when increasing the thermal forcing. Near the onset of convection within the TC, both bulk and wall modes display similar structures to the non-magnetic case. As the thermal forcing is increased these are progressively replaced by a large retrograde vortex occupying the middle of the TC, consistent with the very strong azimuthal flows observed there, see figure \ref{fig:lee1_piv}. This structure however appears at comparably lower thermal forcing, and much higher intensity than in the non-magnetic case and displays off-centre maximum vorticity \cite{aujogue2016_phd}. These results suggest that the axial magnetic field favours the emergence of a large polar vortex, again a plausible, if not directly applicable mechanism for the emergence of polar vortices in the Earth's outer core \cite{hulot2002_nat}.
\subsection{Little Earth Experiment II}
LEE2 is a complete rebuild of LEE1 with much greater control of both the rotation and heating, reduced vibrations and a 10-fold increase in PIV resolution. Its new
vessel focuses on the TC itself so the heater diameter was increased and the outer dome was replaced by a cylinder of diameter \SI{22}{cm} 
with a puck-shaped heater of diameter \SI{15}{cm} and yielding a TC height of $H=$\SI{14.3}{cm}.
Thus, the region outside the TC is now a rather thin rectangular annulus. The outer wall is still cooled by a bath of water at a constant temperature, which imposes a cold temperature along the top and its side walls, i.e. only \SI{3.5}{cm} away from the TC boundary. This creates a much greater source of baroclinicity there than in LEE1 to potentially break away from the TPC, and so attain regimes of higher inertia that would otherwise require higher thermal forcing. 

Inside the new heater, the fluid heat exchanger was replaced by two resistive wires joined together in a spiral shape, and fed with opposite currents to cancel out the total Lorentz force due to the heating current inside the high magnetic field. This made it possible to implement a feedback control loop on the heater to control the temperature difference between any pair of thermistors placed inside the device. Four were located at the top of the heater surface. Two pairs were placed inside and outside of the side wall and the top wall. Measurement inside the vessel provided a more accurate input for the control of the temperature drop within the fluid. 
With this system, the axial temperature difference between the top of the heater and the underside of the top wall could be kept constant for longer periods of time and controlled to $\pm$\SI{0.1}{K}. 
The two thermistor pairs also measured the local heat flux across the outer cylinder top and side walls. 

The new PIV system benefited from a movable vertical LASER plane whose axial position could be remotely controlled during operation, without having to restart the experiment: this made it possible to scan the flow using several PIV planes during the same run. 
All signals were passed to the static frame via an optical slip ring capable of absorbing the bandwidth of the PIV cameras. These were located further away from the bore than in LEE1 to be able to operate at fields up to \SI{12}{T}. These changes made LEE2 considerably heavier, so the new, more powerful motor too had to be placed further away from the magnet and the total height of the device reached \SI{6.5}{m}. LEE2 could also be operated outside M10 in a shorter non-magnetic configuration (\SI{2.5}{m} tall).
\begin{figure}
\includegraphics[width=\textwidth]{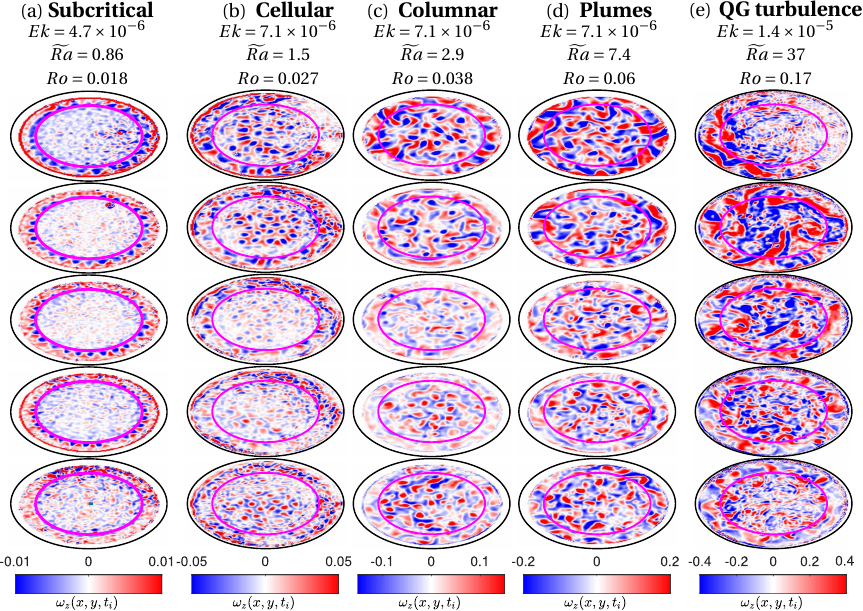}
\caption{Snapshots of axial vorticity obtained from non-simultaneous PIV measurements in LEE2 at $z=H/6,H/3,H/2,2H/3,5H/6$ (from bottom to top)\cite{agrawal2024_gji}. The pink circle marks the location of the TC boundary. Regimes encountered by order of criticality $\Rt$ are similar to those encountered in rotating convection in a cylinder with a solid sidewall. However, the transitions between them take place at lower criticality because of inertia incurred by the baroclinic flow outside the TC. In particular, quasi-geostrophic turbulence appears at $\Rt=37$ \emph{vs.} $\Rt\simeq80$ in Aguirre-Guzman et al.'s work \cite{aguirre-guzman2021_jfm}. The baroclinic flow incurs a local breakup of the TPC visible through structures straddling the TC boundary as low as $\Rt=2.9$ (in the plane at $z=5/6H$).\label{fig:lee2_snapshots}}
\end{figure}

LEE2's high PIV resolution made it possible to map the regimes of convection against the classification established for classical rotating RBC during the 2010's \cite{Julien2012,kunnen2021_jot}. By order of criticality, the flow starts with a regime without convection within the TC, but with a weak toroidal flow in the outer annulus driven by baroclinicity. At slightly higher forcing, the outer flow becomes unstable to baroclinic instabilities and exhibits a ring of co-rotating vortices similar to the "rim instabilities" identified by Aurnou et al. \cite{aurnou2003_epsl}, and visible in figure \ref{fig:lee2_snapshots} (a). Since this region has a constant height, these vortices are not Rossby waves, which they nevertheless resemble. Convection in the TC sets in at approximately the critical Rayleigh number for plane layer rotating RBC and displays the same cellular regime at the onset with the same length scale (figure \ref{fig:lee2_snapshots} (b)). For slightly higher thermal forcing, the cellular network breaks into quasi-geostrophic columns with a symmetric spiral structure of prograde vorticity near the bottom (hot) wall and retrograde vorticity at the top (cold) wall. The structure is surrounded by a sheath of opposite vorticity (figure \ref{fig:lee2_snapshots} (c)). Up to this point, the TC boundary acts essentially as an impermeable wall separating the flow inside and outside the TC with the exception of thin jets induced by baroclinic vortices locally traversing the TC boundary near the top, where baroclinicity is strongest. Further increasing the thermal forcing, convection within the TC enters a plume regime, where columns become three-dimensional, and lose their outer sheath of opposite vorticity. More jets cross the TC near the top and the TC boundary becomes locally permeable over practically the whole height but a small region near the bottom (\ref{fig:lee2_snapshots} (d)). Even higher forcing disrupts the plumes and the flow displays turbulent features with a wide range of scales from small, 3D scales to a small number (two to four) of large vortices. The global balance of force suggests these may still be quasi-geostrophic (QG). However, they appear to be driven by strong baroclinic jets originating outside the TC. In this regime, the TPC is broken over the entire cylinder height but a strong zonal flow persists at its location (\ref{fig:lee2_snapshots} (e)). 

These regimes broadly follow those found in classical rotating RBC \cite{kunnen2021_jot,ecke2023_arfm}. A key feature of the TC geometry is that these regimes become more and more distorted by baroclinically-induced inertia, which progressively loosens the TPC.
The breakup of the TC boundary also enables the heat flux to escape laterally, and thus to bypass the suppression of axial heat transfer by rotation within the TC.   
As a result, the transition between them takes place at significantly lower thermal forcing. A further effect of the strong baroclinicity at the TC boundary is the suppression of wall modes which, unlike in cylindrical vessels appear in none of these regimes. The absence of wall modes and the fact that transitions take place over a narrower range of critical Rayleigh numbers implies that QG turbulence and large-scale vortices \cite{Julien2012,favier2014_pf,guervilly2014_jfm,stellmach2014_prl,aguirre-guzman2021_jfm} are more easily reachable in TC geometries with strong baroclinicity. 

These results suggest that inertia induced by baroclinicity, or otherwise, near the TC boundary in the Earth or other planets may favour the emergence of large structures at lower levels of criticality.  This could potentially explain the large scales observed in the polar regions. Hence, inertia such as that induced by baroclinicity offers a second possible route for the breakup of the TC besides magnetic effects. This mechanism too could favour the meandering of zonal flows in and out of the TC.
LEE2 is poised to examine the competition between these mechanisms in detail with future campaigns in high magnetic fields at LNCMI-G. 
\section{Discussion}
In the end, what does the evolution of experiments modelling rotating convection in planetary interiors over the last 70 years tell us? 
For sure, experiments are hard, cumbersome, and inherently limited by the abysmal gap that separates the relative comfort of the laboratory from realistic planetary conditions: the gravity points in the wrong direction, the fluids are either too viscous or too opaque, the experiments are of the wrong shape, cannot rotate fast enough, reliable measurements are extremely difficult to obtain and the list goes on...
Yet, none of these hurdles have deterred the ingenuity of experimentalists. From Nakagawa's early experiments in cylinders \cite{Nakagawa1955, Nakagawa1957,Nakagawa1959}, to modelling equatorial regions with an annulus \cite{busse1974_jfm} and the solid core with a hockey puck \cite{aurnou2003_epsl} and the more recent mapping velocity fields in vessels with a tangent cylinder \cite{aujogue2016_rsi,aujogue2018_jfm,potherat2023_prl}, they have kept focussing on targeted aspects of planetary interiors: they isolated equatorial and polar regions to obtain Earth-like gravity at least locally, they worked with transparent fluids to reveal the first flow patterns of the outer core \cite{busse1976_sci}, be it at expense of some of the low-$Pr$ convection patterns \cite{Aurnou2018}, they simulated radial gravity in space at the cost of limited ranges of parameters \cite{hart1986_sci,hart1986_jfm,hart1996_3mg}, etc. None of these experiments is a realistic representation of the Earth's outer core, but each of them has revealed key bits of knowledge, which accumulated, and confronted with geophysical data, simulations and mathematical models, have, little by little, shaped our current understanding of planetary cores. The key to their success is that however incomplete, experiments are the best proxy for actual physical processes taking place in planetary interiors, provided the relevant processes can be identified and characterised.

Yet, at a time when numerical simulations have become so accessible and so powerful, and when "realistic" or "Earth-like" simulations are emerging \cite{aubert2023_gji},  it is worth asking whether there is still a place for cumbersome laboratory experiments. The answer lies in the respective merits of both approaches: simulations in extreme regimes will remain time-intensive, at least as costly as experiments and not easily repeatable for the foreseeable future. Unlike experiments, they do offer access to the entire time-dependent field of variables. Unlimited accessibility to an ocean of data is only useful when knowing where to look, and experiments on more targeted processes covering wide parameter ranges are ideally placed to point in the right direction. So are the analysis of real geophysical data and simple mathematical theories. 
Furthermore, even when simulations are perfectly resolved at all scales and all time scales, a very hard condition to meet, the results are still only as good as the mathematical model they are solving. Experiments, on the other hand, are no facsimile: they return the real physics of all the processes they incorporate and so provide a real-life test of the mathematical and numerical models. 
So the answer is yes, experiments are crucial and will continue to improve, just like numerical simulations, and just like the oldest methods in the field: geophysical observation and simple mathematical theories. None of these approaches work anywhere near as well in isolation as when they can draw on the progress and feedback of the other three. 

The main issue faced by experimentalists and theoreticians alike is that of the extrapolation of their results to the Earth or other planets.
Given the limitations of numerics and experiments and the large uncertainties on the actual conditions in planetary interiors, attempting a precise model combining all potentially relevant processes would certainly be in vain. Stevenson finds solace in an elegant way out 
of this conundrum  \cite{Stevenson1985}, citing a similar point made by Urey about the chemical composition of planets \cite{urey1952}: \emph{"What Urey is doing is giving a license to study ill-posed problems. By ill-posed, I mean that the issue can be stated with precision but the knowledge needed to settle the issue is not yet entirely adequate. 
[...] The important lesson that we learn is this: It is not so important that you get the right answer: it is most important that you ask the right question."} In this spirit, the question we may be able to ask may not concern so much the exact nature of the convection in planetary interiors but rather: "What processes arising from the interplay of convection and rotation may we expect to play a role there?". Seven decades of exploring planetary interiors with experiments on rotating convection have certainly provided a wealth of answers to that question but by no means all of them.

So what lies ahead for experiments? The colossal progress of experiments certainly owes to the audacious ideas of their creators, but not only. They also constantly drew on technological and scientific advances from other fields: from the dawn of spacecraft in the 1960s it became possible to escape lab gravity twenty years later \cite{hart1986_sci}. LASERs and digital signal processing led to the advent of optical velocity mapping, ultrasound velocimetry made it possible to measure velocity profiles in opaque metals \cite{brito2001_ef,gillet2007_jfm2,baker2017_ef}.
At the same time, experimental design has incorporated more and more complex aspects of outer core dynamics: cylinders became slanted annuli, then spheres, then spheres with a puck. 
Heating became inhomogeneous, internal \cite{bouillaut2021_pnas, Zaussinger2020, chamberlain1986_gafd} and now magnetic fields are becoming an increasingly important part of the mix.

What is next is certainly found at the crossing of what is currently missing or insufficiently explored in terms of physical processes and what is becoming technologically possible. It is the biased view of the authors that the greatest scope for progress may lie in incorporating MHD effects in the rotating convection experiments. Certainly, the dynamo problem has been the focus of significant experimental endeavours, whose main focus was to find under which conditions a flow could spontaneously generate a magnetic field. The physical processes reproduced in these experiments are largely inspired by potential mechanisms thought to be driving them in planets and stars. {Convection is the best candidate for the Earth and all planets of the Solar system but tidal forces exerted by the Earth 
and precession of the Moon's rotation axis may have driven a past dynamo on the Moon \cite{shea2012_sci,stys2020_jgr}. For Ganymede, Iron snow produced by compositional precession is a possible dynamo mechanism \cite{christensen2015_icarus}. Yet, successful dynamos to date are all mechanically driven: the Karlsruhe dynamo forces a flow through pipes shaped to guarantee the dynamo \cite{stieglitz2001_pf}. In the Riga dynamo, a turbine produces a double helix flow structure \cite{gailitis2000_prl}, and in the Von K\'arman Sodium dynamo (VKS), magnetised counter-rotating impellers generate an unstable von K\'arman flow and produce the dynamo \cite{monchaux2007_prl,berhanu2007_epl}. The upcoming DRESDYN dynamo project in Dresden will spin a \SI{8}{m^3} cylinder filled with Sodium along two axes to produce a precession-driven dynamo \cite{stefani2019_gafd,stefani2024_crphys}. This will potentially be the first dynamo driven by the same process as a geophysical or astrophysical dynamo.} 
Hence, 
the generic character of this question and the magnitude of the challenge involved in building experimental dynamos have pushed aside the question of how actual planetary dynamos work \cite{petrelis2024_crphys}. So understandably, none of the dynamo experiments directly model actual planetary dynamos and the playground of understanding planetary dynamos has been occupied by numericists \cite{glatzmaier1999_nat, Schaeffer2017}. For planets, the question is to find which flow within the outer core produces a magnetic field consistent with observations.
Crucially, the associated Lorentz force due to the magnetic field may drastically alter the answers to that question. In fact, the effects of the Earth's field are often invoked when extrapolating rotating convection experiments to Earth \cite{cardin1994_pepi,aubert2001_pepi,aurnou2003_epsl}. Yet, the Lorentz force is still subject to debate: proponents of the geostrophic scenario argue that it remains much smaller than the Coriolis force \cite{schwaiger2019_gji,aubert2020_gji}, while others argue that a magnetostrophic scenario where both forces are comparable is required to explain the flow patterns inferred from geomagnetic data \cite{dormy2016_jfm,horn2022_prsa}. Here the availability of large electromagnets offers scope for experiments in magnetorotating convection to advance these questions. These possibilities are only starting to be explored: fields up to 18 T in sufficiently large volumes to host laboratory experiments \cite{pugnat2022_ieee} even make it possible to reach extreme regimes where Alfv\'en waves may interact with convection \cite{roberts2000_jfm,lalloz2024_jfm}. Such experiments may also benefit from decades of accumulated knowledge on liquid metal MHD turbulence \cite{alemany1979_jm,messadek2002_jfm,alboussiere1999_etfs,eckert2001_ijhff,potherat2014_jfm,potherat2017_prf,baker2018_prl} and MHD convection tracing back to Nakagawa's first experiments \cite{Nakagawa1955, davoust1999_jfm, Aurnou2001,vogt2018_prf}. Such large magnets combined with measurement techniques developed for transparent electrolytes \cite{andreev2003_pf, andreev2013_jfm, alferenok2013_mst, moudjed2020_ef} now even offer the possibility of mapping velocity and pressure fields in magnetorotating convection at similar $Ek$ to non-MHD experiments but with Elsasser numbers within the range expected for the Earth (between 0.1 and 10) \cite{aujogue2016_rsi,potherat2023_prl}. So perhaps now is the time for Hide's 1958 wisdom to finally come true \cite{hide1958_pta}: \emph{"Now it is clear that extensions in several directions, especially into hydromagnetics, will have to be made if results of any geomagnetic interest are to be obtained."}

Other, non-MHD processes that have been well investigated outside the context of convection in the outer core can be integrated into rotating convection experiments to capture essential elements of the outer core dynamics: solidification and re-melting of the inner core releases or absorbs lighter elements into/from the liquid outer core, and so incurs significant compositional buoyancy \cite{alboussiere2009_nat}. Only one experiment focused on compositional convection in the context of the inner core \cite{cardin1992_grl}. It is sometimes dismissed on the ground that the large Schmidt numbers associated with it justify mimicking convection there with high Prandtl number thermal convection. This analogy may not capture well the crucial conditions at the core boundaries, and even less the interplay between thermal and compositional convection.

Variations of physical properties too may play several roles. In gas giants, they can lead to compressibility effects or to a radial regionalisation of convection forming layers with strongly differentiated convective patterns \cite{leconte2012_aa,gastine2021_icarus,moore2022_jgr,fuentes2024_apjl,horn2024_epsl}. While less significant in the liquid core of Earth-like planets, density variations and the variation of other physical properties may still produce local stratification capable of locally suppressing convection: the presence of such a stratified layer has indeed been theorised near the core-mantle boundary \cite{buffett2010_jgr}.

 The role of the centrifugal force is somewhat less obvious: it remains very small at the scale of planetary cores, but may act at the scale of crucial convection patterns forming there\cite{Horn2021}: tornadoes forming in the Sun offer a reminder not to discard the centrifugal force too quickly in large systems\cite{Luna2015, Kuniyoshi2023}.

Another seven decades will surely tell which of these, or other avenues convection experiments will travel along, but there is little doubt that even then, they will still be advancing in the journey to the outer core of the Earth.
\longthankssection
 We would like to thank John Brothold for his inspiring talk at the 18\textsuperscript{th} SEDI conference at Great Barrington, MA.
 The authors acknowledge the relaxed atmosphere around the time of the XXXIII\textsuperscript{rd} Zappanale in Bad Doberan, Germany, which allowed them to fully focus on this review, and perhaps sipped into some of the wording therein, thusly prompting the question: \emph{"Does humour belong in {[\rm Science]}?"}. 
\bibliographystyle{crunsrt}

\bibliography{ph2024_cras_references}

\end{document}